\documentclass[11pt,letterpaper,twoside,fleqn]{article}
\usepackage{color,graphicx,natbib,lscape,here,geometry,setspace,multirow,enumitem}
\usepackage[dvipsnames]{xcolor}
\usepackage{graphicx}
\usepackage{authblk}
\usepackage{silence}
\usepackage{rotating}
\usepackage{multirow}

\usepackage{booktabs}
\usepackage{color}
\usepackage{setspace}
\usepackage{algorithm2e}
\usepackage{enumitem}
\usepackage{tikz}
\usepackage{placeins}

\usepackage{amsmath}
\usepackage{amssymb}
\usepackage{amsfonts}

\definecolor{nblue}{HTML}{000660}
\usepackage[colorlinks=true,urlcolor=nblue,linkcolor=nblue,citecolor=nblue]{hyperref}
\usepackage{bigstrut}

\usepackage{tabularx}
\usepackage{threeparttable}
\usepackage{dcolumn}

\usepackage{etoolbox} 
\makeatletter
\patchcmd{\BR@backref}{\newblock}{\newblock[}{}{}
\patchcmd{\BR@backref}{\par}{]\par}{}{}
\makeatother

\usepackage{pdflscape}
\usepackage{afterpage}

\usepackage{longtable}
\usepackage{array}
\newcolumntype{C}[1]{>{\centering\arraybackslash}p{#1}}

\usepackage{comment}
\usepackage{mathtools}

\usepackage[hang,flushmargin]{footmisc} 

\usepackage[british]{babel}

\pdfoutput=1
 \usepackage{float}
 \restylefloat{table}

\usepackage[title,titletoc]{appendix}
\makeatletter

\renewenvironment{appendices}{%
    \begin{oldappendices}%
    \renewcommand{\thefigure}{\ifnum \c@section>\z@ \thesection.\fi\@arabic\c@figure}%
    \@addtoreset{figure}{section}%
    \renewcommand{\thetable}{\ifnum \c@section>\z@ \thesection.\fi\@arabic\c@table}%
    \@addtoreset{table}{section}}{%
    \end{oldappendices}%
}\makeatother

\usepackage{titlesec} 
\titleformat{\section}[block]{\large}{\thesection. }{0em}{\MakeUppercase} 
\titleformat{\subsection}[block]{\large}{\thesubsection. }{0em}{\itshape} 
\titleformat{\subsubsection}[block]{\large}{}{0em}{\itshape} 

\let\natbibcitet\citet
\renewcommand\citet{\bibpunct{(}{)}{,}{a}{,}{,}\natbibcitet}
\let\natbibcitep\citep
\renewcommand\citep{\bibpunct{(}{)}{;}{a}{,}{;}\natbibcitep}
\newcommand{\bi}{\begin{itemize}}
\newcommand{\ei}{\end{itemize}}
\newcommand{\be}{\begin{equation}}
\newcommand{\ee}{\end{equation}}
\defcitealias{ieo14}{IEO, 2014}

\makeatletter
\def\ubar#1{\underline{\sbox\tw@{$#1$}\dp\tw@\z@\box\tw@}}
\def\obar#1{\overline{\sbox\tw@{$#1$}\dp\tw@\z@\box\tw@}}
\makeatother

\usepackage{bm}
\usepackage{caption}
\usepackage{subcaption}

\makeatletter\let\p@subfigure\thefigure\makeatother

\floatstyle{plaintop}
\restylefloat{table}

\usepackage{fancyhdr} 
\usepackage{cleveref}
\crefname{chapter}{Chapter}{Chapters}
\crefname{section}{Section}{Sections}
\crefname{subsection}{Section}{Sections}
\crefname{subsubsection}{Section}{Sections}
\crefname{figure}{Figure}{Figures}
\crefname{table}{Table}{Tables}
\crefname{equation}{Equation}{Equations}
\crefname{appendix}{Appendix}{Appendices}
\crefname{appendices}{Appendix}{Appendices}
\crefname{appsec}{Appendix}{Appendices}

\usepackage{catoptions}
\makeatletter

\def\Autoref#1{%
  \begingroup
  \edef\reserved@a{\cpttrimspaces{#1}}%
  \ifcsndefTF{r@#1}{%
    \xaftercsname{\expandafter\testreftype\@fourthoffive}
      {r@\reserved@a}.\\{#1}%
  }{%
    \ref{#1}%
  }%
  \endgroup
}
\def\testreftype#1.#2\\#3{%
  \ifcsndefTF{#1autorefname}{%
    \def\reserved@a##1##2\@nil{%
      \uppercase{\def\ref@name{##1}}%
      \csn@edef{#1autorefname}{\ref@name##2}%
      \autoref{#3}%
    }%
    \reserved@a#1\@nil
  }{%
    \autoref{#3}%
  }%
}
\makeatother

\newcolumntype{d}[1]{D{.}{.}{#1}}
\title{\Large{\textbf{A Bayesian panel VAR model to analyze the impact of climate change on high-income economies}}}
\author{\large{ \MakeUppercase{Florian Huber}$^1$, 
				\MakeUppercase{Tam\'{a}s Krisztin}$^2$\thanks{\noindent\textit{Corresponding author}: T. Krisztin (\href{mailto:krisztin@iiasa.ac.at}{krisztin@iiasa.ac.at}). The authors gratefully acknowledge financial support from the Oesterreichische Nationalbank (Jubilaeumsfond grant no. 17650) and by the Austrian Science Fund (FWF): ZK 35. This paper is a substantially revised version of a manuscript titled ``Dealing with cross-country heterogeneity in panel VARs using finite mixture models.''  \textit{Date}: \today.}\, and 
				\MakeUppercase{Michael Pfarrhofer}$^{1,2}$}\\
\vspace*{-0.5em}\normalsize{$^1$\textit{University of Salzburg}\\
							$^2$\textit{International Institute for Applied Systems Analysis (IIASA)}}}
\date{}


\def\equationautorefname~#1\null{%
  Eq.~(#1)\null
}
\def\equationautorefname~#1\null{
Eq.~(#1)\null
}

\usepackage[utf8, latin1]{inputenc}                 

\graphicspath{{Results/}}

\begin{document}
\maketitle\thispagestyle{empty}\normalsize\vspace*{-2em}\small

\begin{center}
\begin{minipage}{0.8\textwidth}
\noindent\small In this paper, we assess the impact of climate shocks on futures markets for agricultural commodities and a set of macroeconomic quantities for multiple high-income economies. To capture relations among countries, markets, and climate shocks, this paper proposes parsimonious methods to estimate high-dimensional panel VARs. We assume that coefficients associated with domestic lagged endogenous variables arise from a Gaussian mixture model while further parsimony is achieved using suitable global-local shrinkage priors on several regions of the parameter space. Our results point towards pronounced global reactions of key macroeconomic quantities to climate shocks. Moreover, the empirical findings highlight substantial linkages between regionally located climate shifts and global commodity markets.
\end{minipage}
\end{center}

\normalsize\renewcommand{\thepage}{\arabic{page}}
\newpage

\section{Introduction}
A projected increase in extreme climate events and an increasingly interdependent food supply chain pose a threat to global food security. Increasing trade flows and the rising complexity of economic networks lead to higher vulnerability to systemic disruptions \citep{Puma2015}. Isolating the effects of climate related production shocks on agricultural commodity markets, food prices, and the globalized economy is thus of special interest for policy makers and the wider public in general.

Global commodity markets play a crucial role in establishing a relationship between agricultural production and the economy. Due to increased demand and limited production capabilities, volatilities associated with agricultural prices are expected to rise over the next decades \citep{FAO2017,IFPRI2008}. Among the key drivers of increasing volatility in related prices are exogenous weather and production shocks, but also influences from other economic sectors \citep[e.g. demand, energy market and exchange rate market shocks, see, for instance,][]{Gilbert2010,Serra2011,Nazlioglu2011a,Nazlioglu2012}. Relatedly, changes in fiscal and monetary policies affect food price volatility \citep{Akram2009,Baffes2010}. 

Besides an overall trend of increased volatilities in commodity prices, linkages across agricultural and energy markets increased in recent years, for instance due to the rising importance of biofuels. This lead to intensified competition for food production resources \citep{Harri2009, Saghaian2010, Havlik2011,Nazlioglu2011,Enders2014}. Such linkages are expected to strengthen further as a consequence of climate change. This in turn calls for research on the effects of climate shocks and their respective impact on real and financial economic sectors across economies, with a special focus on food prices, and feedback and spillover effects between countries \citep{Gilbert2010,Jebabli2014}.

International trade is generally perceived as an important mitigation mechanism of economic fluctuations, but also as a potential source of volatility \citep{Huang2011,Gaupp2017,Sandstrom2018}. Addressing cross-country interdependencies on the macro-level is crucial, especially when aiming to capture the international effects of climate change in highly globalized markets. The vast majority of the literature on volatility transmissions from climate change on agricultural markets and food security, however, neglects such notions, but instead focuses solely on the nexus between global commodity markets and climate shocks \citep{Harri2009,Nazlioglu2011,Enders2014,Garcia2015}. Such approaches disregard the fact that global commodity prices do not capture country-specific movements in food prices, which might depart considerably from global movements in prices. Apart from considering global commodity markets, studies focusing on the impact of climate-related shocks on country-specific macroeconomic and commodity market-related quantities typically fail to take potential spillovers and feedback effects from trade, exchange rates and other global factors into account \citep[see, e.g.,][]{Guerrero2017,VanHuellen2018}.
 
Integrated global supply chains and international trade call for a multi-country setup that takes into account cross-country spillovers and feedback effects between economies. Popular large-scale macroeconometric models, however, are typically heavily parameterized. This often translates to imprecise inference, rendering policy relevant conclusion difficult. From a methodological point of view, the main contribution of this paper is to propose a parsimonious yet flexible approach to estimate panel vector autoregressive (PVAR) models. We combine the literature on Bayesian PVAR models \citep[see][]{canova2004forecasting, canova2009estimating, koop2016model, korobilis2016prior} with the literature on finite mixture models \citep[see][]{allenby1998heterogeneity, lenk2000bayesian, fruhwirth2004bayesian, frohwirth2008model, malsiner2016model}. Dependency structures across economies are pushed to zero by means of global-local shrinkage priors in the spirit of \cite{griffin2010inference} and \cite{Huber2017}. To account for international comovement of volatilities that vary over time, we assume that the errors of the system feature a factor stochastic volatility structure. This provides a parsimonious representation for a high-dimensional time-varying variance-covariance matrix \citep[see][]{kastner2017sparse,Kastner2016}.

The empirical contribution deals with the highly policy-relevant question of how climate shocks impact country-specific macroeconomic fundamentals. To control for global movements in commodity markets, the baseline framework is augmented by a separate global model. This implies that we jointly model 17 OECD economies, where each country-specific model features several macroeconomic indicators as well as food prices. We specifically focus on high-income developed economies to demonstrate the impact of climate change on the macroeconomy of countries whose agricultural production constitutes only a small part of output. The global model consists of commodity futures data from the United States, rendering our model a mixed frequency panel VAR akin to the model proposed in \cite{georgiadis2015examining}. In our modeling framework climate shocks -- assessed in terms of agricultural production under risk of drought in eleven global regions -- are treated as strictly exogenous. Our results demonstrate that climate shocks have a substantial impact on short-term interest rates and inflation, and to a lesser degree on output and exchange rates. Significant impacts are present, even if the regions hit by climate shocks are neither in the countries themselves, nor among major trading partners, likely due to the integration of global financial markets. 

The paper is structured as follows. Section \ref{sec: econometrics} introduces the general econometric framework and specifies the adopted prior setup. We proceed by introducing the novel dataset in Section \ref{sec:dataandmodel}. This section also includes further details on the model specification. The empirical findings are discussed in Section \ref{sec: application}. The last section summarizes and concludes the paper.

\section{Econometric framework}\label{sec: econometrics}
In this section we discuss the PVAR model along important specification issues in Subsection \ref{subsec:PVAR}, while the remainder of the section is devoted to dealing with these issues using flexible Bayesian shrinkage priors. Before proceeding to the model, it is convenient to introduce generic notation. In what follows, capitalized letters without a time index refer to full-data matrices, i.e. $\bm{Y} =  (\bm{y}_1, \dots, \bm{y}_T)'$, unless otherwise noted.  The notation $[\bm{Y}]_{j \bullet}$ selects the $j$th row of the matrix $\bm{Y}$ while $[\bm{Y}]_{\bullet j}$ selects the $j$th column of the matrix concerned. In addition, we let $\bm{y}_{-i, t}$ denote the vector $\bm{y}_t$ with the $i$th subvector excluded, i.e. $\bm{y}_{-i t}=(\bm{y}'_{1t},\dots, \bm{y}'_{i-1t},\bm{y}'_{i+1t},\dots,\bm{y}'_{Nt})' $. Finally, we let $\bm{\bullet}$ be a generic notation  that indicates conditioning on all remaining coefficients in the model as well as the data. 

\subsection{The panel vector autoregressive model}\label{subsec:PVAR}
In this paper, we aim to model  a set of $M$ macroeconomic and financial indicators across a set of $N$ countries. For each country, the domestic quantities are stored in an $M$-dimensional vector $\bm{y}_{it}$ for $i=1,\dots,N$ and $t=1,\dots,T$, and consequently stacked in a vector $\bm{y}_t=(\bm{y}'_{1t},\dots, \bm{y}'_{Nt})'$ of dimension $K=MN$. We assume that $\bm{y}_{it}$ follows
\begin{equation}
\bm{y}_{it} = \bm{\beta}_i+\bm{A}_{i1} \bm{y}_{it-1}+\dots+\bm{A}_{iP} \bm{y}_{it-P} +\bm{B}_{i1}\bm{y}_{-i, t-1}+\dots+\bm{B}_{iP}\bm{y}_{-i, t-P}
+\bm{\varepsilon}_{it},\label{eq:PVAR}
\end{equation}
where  $\bm{\beta}_i$ is an $M$-dimensional intercept vector and $\bm{A}_{ij}~(j=1,\dots,P)$ denotes a set of $M \times M$-dimensional coefficient matrices associated with the $P$ lags of $\bm{y}_{it}$. In what follows, we label these parameters the domestic VAR coefficients. The impact of other countries' lagged dependent variables $\bm{y}_{-i, t-p}$ is measured through the matrices $\bm{B}_{ij}$, which are of dimension $M \times (N-1)M$. Finally, $\bm{\varepsilon}_{it} \sim \mathcal{N}(\bm{0}_M, \bm{\Sigma}_{it})$ is a Gaussian vector white noise process with a time-varying variance-covariance matrix $\bm{\Sigma}_{it}$.

Equation (\ref{eq:PVAR}) can be cast in the usual regression form,
\begin{equation}
\bm{y}_{it} = \bm{C}_{i} \bm{x}_{it}+ \bm{B}_i \bm{x}_{-i, t}+ \bm{\varepsilon}_{it},\label{eq:PVARstacked}
\end{equation}
with $\bm{x}_{it}= (1,\bm{y}'_{it-1},\dots, \bm{y}'_{it-P})', \bm{C}_i = (\bm{\beta}_i, \bm{A}_{i1},\dots,\bm{A}_{iP}), \bm{x}_{-i, t}=(\bm{y}'_{-i, t-1}, \dots, \bm{y}'_{-i, t-P})' $ and $\bm{B}_i = (\bm{B}_{i1}, \dots, \bm{B}_{iP})'$. The matrix $\bm{B}_i$ establishes DIs between countries $i$ and $j$.  In the  literature on PVAR models \citep[see][for a recent survey]{canova2013panel}, an important modeling decision is whether to set certain sub-matrices of $\bm{B}_i$ to zero, shutting off dynamic relations between country pairs. An extreme version of the model would set the whole matrix $\bm{B}_i$ to zero, ruling out lagged relations between country $i$ and the remaining economies.

Up to this point, we remained silent on assumptions regarding error covariances across countries. Here, we stack the country-specific errors $\bm{\varepsilon}_{it}$ in a $K$-dimensional vector $\bm{\varepsilon}_t = (\bm{\varepsilon}_{1t}',\hdots,\bm{\varepsilon}_{Nt}')'$ and assume that
\begin{equation}
\bm{\varepsilon}_t\sim \mathcal{N}(\bm{0}, \bm{\Sigma}_t),\label{eq:errorterm}
\end{equation} 
where $\bm{\Sigma}_t$ is a full $K \times K$-dimensional variance covariance matrix. 

High-dimensional PVAR models, such as the one proposed in Eqs. (\ref{eq:PVAR}) to (\ref{eq:errorterm}), are highly parameterized and model uncertainty is pervasive. Three important dimensions of model uncertainty have been identified by the literature. The first one is concerned with modeling contemporaneous relations across the shocks in the system (called static interdependencies, SIs) while the second dimension centers on the question whether coefficients associated with lagged domestic variables are homogeneous across countries (labeled homogeneity restrictions). If such domestic coefficients are similar, so-called homogeneity restrictions might be imposed, effectively introducing the same set of coefficients for several countries and therefore reducing the number of free parameters.  The final dimension deals with the question on whether to allow  for lagged dependencies between countries (labeled dynamic interdependencies, DIs).

Considering the recent literature on model specification and selection in PVAR models reveals two commonly used approaches to deal with the aforementioned issues. The first strand of the literature suggests applying shrinkage priors to stochastically select an appropriate model specification \citep[see][]{koop2016model, korobilis2016prior}. In light of the large number of potential restrictions, however, mixing issues typically arise, leading to weak convergence properties of existing algorithms \citep{bhattacharya2015dirichlet}. The second strand considers additional restrictions that reduce the dimension of the parameter space. For instance, \cite{canova2009estimating} assume that the (time-varying) coefficients of the PVAR model feature a factor structure. This translates into statistical and computational gains since the dimension of the state space is substantially reduced. Another prominent example are global VAR models \citep{Pesaran2004,Dees2007a, feldkircher2016international, cuaresma2016forecasting, huber2016density} that introduce parametric restrictions on the coefficients associated with other countries' endogenous variables.

\subsection{Dealing with static interdependencies}
In this section, we start with discussing how to tackle the first dimension of model uncertainty. SIs are introduced by using  a factor stochastic volatility structure \citep{pitt1999time, aguilar2000bayesian,Kastner2016} on $\bm{\Sigma}_t$, 
\begin{equation}
\bm{\Sigma}_t = \bm{L} \bm{H}_t  \bm{L}'+\bm{\Omega}_t. \label{eq: factorSV}
\end{equation}
$\bm{L}$  is a $K \times q$ matrix  of factor loadings (with $q \ll K$),  $\bm{H}_t=\text{diag}(e^{h_{1t}}, \dots, e^{h_{qt}})$ is a diagonal matrix containing the variances of a set of $q$ common factors  $\bm{f}_t \sim \mathcal{N}(\bm{0}, \bm{H}_t)$, and $\bm{\Omega}_t= \text{diag}(e^{\omega_{1t}}, \dots, e^{\omega_{K t}})$ is a diagonal variance-covariance matrix of idiosyncratic shocks $\bm{\eta}_t \sim \mathcal{N}(\bm{0}, \bm{\Omega}_t)$. 

An equivalent representation of \autoref{eq: factorSV} is the regression form,
\begin{equation*}
\bm{\varepsilon}_t = \bm{L} \bm{f}_t + \bm{\eta}_t.
\end{equation*}
Hereby, the key feature from a computational point of view is that conditional on $\bm{L} \bm{f}_t $, the PVAR reduces to a system of unrelated regression models. This leads to substantial computational gains relative to full system estimation \citep[see][for more details and an efficient algorithm]{kastner2017sparse}.

 We assume that the (log) of the main diagonal elements of $\bm{H}_t$ and $\bm{\Omega}_t$ follow independent AR(1) processes,
\begin{align}
h_{jt}  &= \rho_{hj} h_{jt-1} + \sigma_{hj} \zeta_{hj, t}, \text{ for } i=1,\dots,q, \label{eq:hsv}\\
\omega_{jt} &= \phi_{\omega j} + \rho_{\omega j} (\omega_{jt-1}-\phi_{\omega j}) + \sigma_{\omega j} \zeta_{\omega j, t}, \text{ for } j=1,\dots,K.
\end{align}
We let $ \phi_{\omega j}$ denote the unconditional mean of the log-volatility, $\rho_{sj}$ the autoregressive parameter and $\sigma^2_{sj}$ the process innovation variance for $s \in \{h, \omega\}$.  Moreover, $\zeta_{sj, t} \sim \mathcal{N}(0,1)$ is a serially uncorrelated white noise shock. To identify the unconditional scaling of the factors, \autoref{eq:hsv} is assumed to have zero mean \citep{Kastner2016}. 

Notice that opposed to ${K (K+1)}/{2}$ total parameters in the case of an unrestricted $\bm{\Sigma}_t$, the structure in \autoref{eq: factorSV} implies that we only have to estimate $(K+1)q +K$ coefficients, a  substantial reduction relative to an unrestricted variance-covariance matrix if $q$ is small. One important consequence of \autoref{eq: factorSV} is that the covariance structure of the errors is driven by relatively few latent factors that summarize the joint dynamics of $\bm{\varepsilon}_t$.

\subsection{Dealing with parameter homogeneity}
It is worth noting that the total number of parameters of the PVAR model outlined in the previous section is $K (p K+1)+(K+1)q +K$, and thus rises rapidly with $K$ (and implicitly with $M$ and $N$). Since typical macroeconomic datasets include time series with only a few hundred observations, some form of regularization is needed.  To cope with this issue, the Bayesian literature suggested various means of achieving parsimony in the PVAR framework. One strand of the literature uses shrinkage priors on several parts of the parameter space \citep{koop2016model, korobilis2016prior, koop2018forecasting}. This approach conceptually treats the PVAR as a large VAR with asymmetric shrinkage on the different coefficients in $\bm{A}_i, \bm{B}_i$ and the free elements in $\bm{\Sigma}_t$. Another strand \citep{canova2004forecasting, canova2009estimating, jarocinski2010responses} exploits the observation that countries do not differ much in terms of their domestic macroeconomic dynamics, implying that the matrices $\bm{A}_i$ tend to be similar across countries. This literature often pools information across countries by shrinking towards a common mean of $\bm{A}_i$, but neglects dynamic or static interdependencies.

We deal with the second pillar of model uncertainty (parameter homogeneity across countries) by  assuming that the domestic  coefficients $\bm{c}_i = \text{vec}\{\bm{C}_i\}$ arise from a  $G$-component mixture of Gaussians distribution. A variant of this model has been proposed in the marketing literature \citep{allenby1998heterogeneity, lenk2000bayesian, fruhwirth2004bayesian} and is commonly referred to as the heterogeneity model.   In the present framework, the mixture distribution for $\bm{c}_i$ is given by,
\begin{equation}
p(\bm{c}_i|\bm{w}, \bm{\mu}_1, \dots, \bm{\mu}_G, \bm{V}) = \sum_{g=1}^G w_g f_\mathcal{N}(\bm{c}_i | \bm{\mu}_g, \bm{V}).\label{eq: mixtures}
\end{equation}
Hereby, $\bm{w}=(w_1, \dots, w_G)'$ is a vector of component weights that satisfy $\sum_{g=1}^G w_g =1$ and $w_g \ge 0$. Additionally, $f_\mathcal{N}$ is the density of the multivariate Gaussian distribution, $\bm{\mu}_g$ is an $m = M (Mp+1)$-dimensional component-specific mean vector, and $\bm{V}$ is a common variance-covariance matrix.   This specification assumes that coefficients of countries within a given country group tend to be similar, with potential deviations from  $\bm{\mu}_g$ driven by $\bm{V}$. 

To estimate the mixture model, we introduce a set of $N$ binary indicators $\delta_i$ that allow to state \autoref{eq: mixtures} as,
\begin{equation*}
p(\bm{c}_i|\delta_i = g, \bm{\mu}_g, \bm{V}) = f_\mathcal{N}(\bm{c}_i | \bm{\mu}_g, \bm{V}),
\end{equation*}
with $\Pr(\delta_i = g) = w_g$. In what follows we  exploit this auxiliary representation for estimation of the mixture model. Notice that  ergodic averages of the posterior draws of $\delta_i$ can be used to obtain the probability that country $i$ is located within a specific country group.

On the main diagonal elements of $\bm{V}$, we apply a set of independent inverted Gamma priors,
\begin{equation*}
v_j \sim \mathcal{G}^{-1}(w_0, w_1), \text{ for } j=1,\dots,m,
\end{equation*}
with the hyperparameters $w_0$ and $w_1$ typically set to small values, i.e. $w_0=w_1=0.01$. This leads  to a weakly informative prior on the common variances.  

Another key assumption is that each mixture component again comes from a common distribution,
\begin{equation*}
\bm{\mu}_g|\bm{\mu}_0, \bm{Q}_0 \sim \mathcal{N}(\bm{\mu}_0, \bm{Q}_0) \text{ for } g=1,\dots,G.
\end{equation*}
We let $\bm{\mu}_0$ denote a common mean and $\bm{Q}_0$ is a diagonal  variance-covariance matrix that can be decomposed as
\begin{equation*}
\bm{Q}_0 = \bm{\Lambda} \bm{R}_0 \bm{\Lambda},
\end{equation*}
where the matrix $\bm{\Lambda} =\text{diag}(\sqrt{\lambda_1}, \dots, \sqrt{\lambda_m})$ contains the standard deviations and $ \bm{R}_0=\text{diag}(R_1^2, \dots, R^2_m)$ constitutes an additional scaling matrix with $R_j^2$ denoting the range of $\bm{c}=(\bm{c}_1, \dots, \bm{c}_N)$ along the $j$th dimension \citep[see][]{malsiner2016model}.

\subsection*{Selecting cluster-relevant quantities}
To select the driving forces behind  cluster allocation, we follow \cite{yau2011hierarchical} and consider the standardized distance between cluster centers for a given element  $j$ of  $\bm{\mu}_i$ for clusters $g$ and $s$,
\begin{equation*}
\frac{(\mu_{g j} - \mu_{s j})}{\sqrt{2 R^2_j}} \sim \mathcal{N}(0, \lambda_j)~\text{ for } j=1,\dots,m.
\end{equation*}
By specifying a suitable mixing density on $\lambda_j$, we can flexibly shrink the distance between cluster centers to zero and thus are able to identify cluster relevant variables.  As an example, consider a situation where the conditional mean of output growth strongly differs across countries while the remaining quantities (i.e. the coefficients associated with the lags of $\bm{y}_{it}$) display only minor differences. In such a situation, a shrinkage prior would strongly pull the cluster centers together  for elements in $\bm{\mu}$ not related to the intercept, while at the same time allowing for large differences between the cluster means for the intercept terms.

Following \cite{malsiner2016model}, we introduce a Gamma prior on $\lambda_j$, leading to a variant of the Normal-Gamma (NG) shrinkage prior \citep{griffin2010inference}. More specifically, we set
\begin{equation*}
\lambda_j \sim \mathcal{G}(\nu_1, \nu_2),
\end{equation*}
where $\nu_1$ and $\nu_2$ are hyperparameter specified by the researcher.  Notice that if $\nu_1=1$, we obtain the Bayesian Lasso \citep{park2008bayesian} used in \cite{yau2011hierarchical}. The NG prior improves upon the Lasso by featuring a marginal prior that possesses heavier tails than the Laplace distribution. In fact, the marginal prior of the proposed specification is available in closed form \citep{fruhwirth2011label},
\begin{equation}
p(\mu_{1j}, \dots, \mu_{Gj}|\bm{\mu}_0) = \frac{\nu_2^{\nu_1}}{(2\pi)^{G/2} \Gamma(\nu_1)} 2 K_{p_G}(\sqrt{d_j e_j}) \left(\frac{e_j}{d_j}\right)^{p_G/2}, \label{eq:margprior}
\end{equation}
with $d_j=2 \nu_2, p_G= \nu_j - G/2$, $e_j = \sum_{g=1}^G (\mu_{gj}-\mu_{0j})^2/R_j^2$ and $\Gamma(\star)$ is the Gamma function. In addition,  $K_{\alpha}(\star)$ represents the modified Bessel function of the second kind and $\mu_{0j}$ denotes the $j$th element of $\bm{\mu}_0$. \cite{griffin2010inference} show that the excess kurtosis of the NG prior is given by $3/\nu_1$ and thus rises with smaller values of $\nu_1$. If $\nu_1$ is close to zero, more mass is placed on zero while at the same time maintaining heavy tails of the marginal prior. In the applications, we specify $\nu_1=\nu_2=1/2$ to strongly push the standardized distance between cluster centers to zero.

The prior on $\bm{\mu}_0 \sim \mathcal{N}(\bm{m}_0, \bm{M}_0)$  is  improper with $\bm{m}_0$ denoting the median over the columns of $\bm{c}$ and  $\bm{M}^{-1}_0 = \bm{0}$. Here, one alternative would be to use a Minnesota prior \citep{Doan1984} at the top level of the hierarchy, assuming that $\bm{\mu}_0$ again features a normally distributed prior centered on a multivariate random walk with a known prior variance-covariance matrix. For several datasets, however, we found that this choice only exerts a minor impact on the results.

\subsection*{Choosing the number of mixture components}
To endogenously select the number of components $G$, we follow \cite{malsiner2016model} and introduce a symmetric Dirichlet prior on the mixture component weights $\bm{w}$,
\begin{equation*}
\bm{w} \sim Dir(p_0, \dots, p_0),
\end{equation*}
where $p_0$ denotes the intensity parameter of the Dirichlet distribution.  In the framework of overfitting mixture models (i.e. models that set $G$ greater than the true number of clusters,  $G^{true}$), the parameter $p_0$ plays an important role in shaping the way the posterior distribution treats redundant mixture components.\footnote{For a discussion, see \cite{fruhwirth2006finite} and \cite{rousseau2011asymptotic}} 

In what follows, we place another Gamma prior on $p_0$. Following \cite{ishwaran2001bayesian} and \cite{malsiner2016model}, we choose a Gamma prior with expectation $E(p_0)=1/G$, 
\begin{equation*}
p_0 \sim \mathcal{G}(c_0,c_0 G).
\end{equation*}
Hereby, we let $c_0$ be a hyperparameter that controls the variance of the prior $1/(c_0 G^2)$. This choice handles irrelevant mixture components by shrinking the associated weights to zero and empties superfluous components. Consistent with simulation evidence provided in \cite{malsiner2016model}, we set $c_0=10$.

\subsection{Dealing with dynamic interdependencies}
To decide on whether DIs, the third aspect of model uncertainty in PVARs, for a given country $i$ are present, we use a NG shrinkage prior similar to the one discussed above.  While the prior on $\bm{\mu}_0$ introduces local shrinkage parameters that push the differences between cluster centers towards zero, the standard implementation of the NG prior combines local shrinkage parameters  with a global shrinkage factor that pulls all coefficients  concerned to zero.

To illustrate the problem of selecting DIs, we partition the matrices $\bm{B}_{ip}$ for $p=1,\dots,P$ and stack them to obtain
\begin{equation*}
\bm{B}_p =\begin{pmatrix}
\bm{B}_{1p}\\
\bm{B}_{2p}\\
\vdots\\
\bm{B}_{Np}
\end{pmatrix}=
\begin{pmatrix}
\bm{B}_{12, p}&  \bm{B}_{13, p}&\dots &\bm{B}_{1N, p}\\
\bm{B}_{21,p} &\bm{B}_{23,p}& \ddots &\vdots \\
\vdots& \ddots & \vdots &\bm{B}_{N-1 N,p}\\
\bm{B}_{N1,p}  & \dots & \bm{B}_{N N-2,p}&\bm{B}_{N N-1,p}
\end{pmatrix},
\end{equation*}
where the submatrix $\bm{B}_{ij, p}$ measures the DIs between countries $i$ and $j$ for lag $p$.  Model specification boils down to deciding whether a given $\bm{B}_{ij, p}$ equals zero, ruling out DIs between countries $i$ and $j$. \cite{koop2016model} use a stochastic search variable selection (SSVS) prior that is based on a set of auxiliary indicators that determine whether different sub-matrices  of $\bm{B}_p$ are pushed to zero.  While this approach is conceptually straightforward to implement,  a high dimensional model space needs to be explored. Using MCMC techniques helps to circumvent this issue by performing a stochastic model specification search that only explores a fraction of the full model space. However, in large dimensions, the possible number of DI restrictions  is huge, even for a moderate number of countries included. In that case, even SSVS priors manage to exploit only a tiny fraction of the model space, leading to weak convergence \citep{bhattacharya2015dirichlet}.

In this paper, we  assume that each element of $\text{vec}(\bm{B}_i)$, labeled $b_{ij}$, features a normally distributed prior,
\begin{equation}
b_{ij}| \tau_{ij}, \xi_i \sim \mathcal{N}\left(0, \frac{2 \tau^2_{ij}}{\xi_i}\right),\quad \tau^2_{ij} \sim \mathcal{G}(\vartheta_i, \vartheta_i), \quad \xi_i \sim \mathcal{G}(\mathfrak{c}_0, \mathfrak{c}_1), \label{eq:NG_int}
\end{equation}
for $j=1,\dots, k=P M^2 (N-1)$ and $i=1,\dots,N$. $\xi_i$ denotes a country-specific global scaling parameter that pushes all elements in $\bm{B}_i$ (or equivalently $\bm{B}_{ip}$ for all $p$) to zero, shutting off DIs between a given country and all remaining countries, if necessary. Overall shrinkage is driven by the hyperparameters $\mathfrak{c}_0, \mathfrak{c}_1$, with small values translating into heavy shrinkage.  

Since shutting of all DIs within a given country would be overly restrictive, we introduce a set of local scaling parameters $\tau^2_{ij}$. The local scaling parameters  allow for  non-zero $b_{ij}$'s, even in the presence of strong global shrinkage due to a heavy tailed marginal prior (see \autoref{eq:margprior}), with excess kurtosis depending on $\vartheta_i$. This enables flexible selection of restrictions  of the form whether country $i$'s  output depends on country $c$'s lagged output while turning off dependencies between output in country $i$ and, for instance, lagged interest rates in country $c$. We set $\mathfrak{c}_0=\mathfrak{c}_1=0.01$ and $\vartheta_i=0.1$. Both hyperparameter values are based on evidence in \cite{Huber2017}, who integrate out $\vartheta_i$ in a Bayesian fashion and find values between $0.1$ to $0.3$, depending on the size of the model involved.

\subsection{Priors for the factor stochastic volatility specification}
For the remaining coefficients we utilize the prior setup proposed in \citet{Kastner2016}. In particular, we use a row-wise NG shrinkage prior on the factor loadings in $\bm{L}$, with its elements denoted by $l_{ij}$.\footnote{For estimation, we rely on the efficient implementation provided in the \texttt{R}-package \texttt{factorstochvol} \citep{factorstochvol}. Further details on potential variants of the Normal-Gamma shrinkage prior such as column-wise shrinkage may be obtained in \citet{Huber2017}.} We specify the prior
\begin{equation*}
l_{ij}|\varphi_{ij},\zeta_i \sim \mathcal{N}\left(0,\frac{2\varphi_{ij}^2}{\zeta_i}\right), \quad
\varphi_{ij}^2\sim\mathcal{G}(\vartheta_l,\vartheta_l), \quad
\zeta_i\sim\mathcal{G}(\mathfrak{e}_0,\mathfrak{e}_1).
\end{equation*}
The hyperparameters are set to $\vartheta_l=0.1$ and $\mathfrak{e}_0=\mathfrak{e}_1=1$. On the parameters of the state equations for the log-volatility processes, we use a normally distributed prior on the unconditional mean $\mu_{\omega j} \sim \mathcal{N}(0, 10)$ for all $j$, a Gamma prior on the process innovation variances $\sigma^2_{sj} \sim \mathcal{G}(1/2, 1/2)$, and a Beta prior on the (transformed) autoregressive parameter ${(\rho_{sj}+1)}/{2}\sim \mathcal{B}(10,3)$ for all $s,j$. Using different hyperparameter values or estimating the model with weakly informative independent Gaussian priors on the factor loadings has negligible consequences for the results in \cref{sec: application}.

\subsection{Identification issues}\label{sec:ident}
The model described above is econometrically not identified, with identification issues stemming from  two sources. First, the factor model in \autoref{eq: factorSV} is not identified unless suitable restrictions are introduced. Here, we employ an automatic restriction search approach implemented in \citet{factorstochvol}. The second source arises from the well known label switching problem.\footnote{For a discussion, see \cite{fruhwirth2006finite}.} This issue  comes from the invariance of the mixture likelihood function in \autoref{eq: mixtures} with respect to relabeling the components,
\begin{align*}
p(\bm{c}_i|\bm{w}, \bm{\mu}_1, \dots, \bm{\mu}_G, \bm{V})& = \sum_{g=1}^G w_g f_\mathcal{N}(\bm{c}_i | \bm{\mu}_g, \bm{V})\\
&=\sum_{g=1}^G w_{\varrho(g)} f_\mathcal{N}(\bm{c}_i | \bm{\mu}_{\varrho(g)}, \bm{V}),\nonumber
\end{align*}
with $\varrho$ indicating a random permutation of $\{1, \dots, G\}$.  We obtain identification by applying the random permutation sampler outlined in \cite{fruhwirth2001markov}  and then perform ex-post identification of the model. In our case, and since $N$ is typically a moderate number of countries, we can easily identify  different country groups via economic reasoning. In the empirical application, for instance, we introduce an ordering constraint on the size of the cluster components. Furthermore, notice that if interest centers exclusively on functionals of the coefficients in \autoref{eq:PVAR}, such as impulse response functions or predictive densities, obtaining explicit identification is not necessary. However, it is worth emphasizing that if unbalanced label switching takes place (i.e. the posterior simulator jumps only between a small number of the $G!$ potential modes), inference could be distorted. Using the random permutation sampler in that situation thus leads to balanced label switching, ensuring that the algorithm visits all modes.

This completes the prior setup of our modeling approach. To obtain posterior distributions for all parameters, we propose a Markov chain Monte Carlo (MCMC) algorithm that consists of several blocks. We briefly summarize the algorithm with all full conditional posterior distributions in \cref{sec:posteriorsim}. Moreover, for an illustration of the merits of our approach using synthetic data, see \cref{sec: simulation}.

\section{Data and model specification}\label{sec:dataandmodel}
In this section, we present a novel dataset to assess the impact of climate shocks on futures markets for agricultural commodities and key macroeconomic quantities across a set of OECD economies. Moreover, we discuss a further modification of the general model framework described in \cref{sec: econometrics}.

\subsection{Data and descriptives}
Our dataset contains monthly observations for 17 OECD member countries, comprised of 13 European member states, Canada, Israel, South Africa, and the US. Jointly our sample covers a majority of high income economies (as defined by the World Bank), accounting for 77 percent of the population and 82 percent of output in terms of nominal GDP in US dollars.

The sample includes $215$ monthly observations per country, ranging from January 2000 to November 2017. This encompasses multiple major drought shocks in Europe and the United States, as well as the food crisis of 2007/08, where the prices of agricultural commodities experienced a sharp peak. The country-specific time series include the total consumer price index (CPI), short-term interest rates (ir), the total value of industrial production (prod), and the real effective exchange rate to the US dollar (XR2). Movements in food prices are measured through the ratio of the consumer price index for food products to the total consumer price index (CPF). 

In order to capture the role of the futures market, our dataset includes monthly observations on continuous Chicago Mercantile Exchange (CME) and Intercontinental Exchange (ICE) future prices with a two-month forward contract for eleven commodities. While the data mainly covers the US futures market, there is strong evidence suggesting that commodity markets are highly integrated, similar to financial markets \citep[for example][]{Nazlioglu2011, Huber2016,Cashin2017a}. This has been emphasized in the literature especially since the commodity boom in 2004/05. Global shocks to agricultural production are thus expected to be reflected in the US futures market \citep{Nazlioglu2011a,Headey2011}. Our main crop production indicators are rice, corn, cotton, wheat and soybean futures. Additionally, to assess the role of the livestock sector, we include hogs, feeder and live cattle futures.\footnote{In all three cases of livestock futures the actual traded good is the slaughtered, packaged, and frozen meat of the animals.}  Feeder cattle are freshly weaned calves, whereas live cattle are fully grown animals. The interplay of these two livestock futures allows us to gauge how shocks to feed supply play out in the markets. Moreover, the interlinkages of energy and agricultural markets -- specifically the oil and ethanol markets -- is well-established in literature \citep{Nazlioglu2011a,Nazlioglu2011,Lucotte2016}. To measure the responses of the energy and biofuel sectors to climate shocks, we include crude oil, gasoil and ethanol futures in  our model. A full list of indicators, as well as our selection of countries, is presented in \cref{sec: app_regdat}.

Finding an indicator which accurately reflects the presence or absence of climate shocks in agricultural production regions at a national level poses a serious challenge. National level average of climate indicators such as precipitation or temperature  fail to take into account the regional variation and localized impact of climate change \citep{Harari2018,Burke2019}. To alleviate this issue, we rely on a spatially explicit dataset to proxy climate change impacts. This allows us to take the localized nature of climate change into account. This data is subsequently aggregated to the supra-national level, weighted by a high resolution dataset of agricultural production. The advantage of this approach is that we explicitly capture  localized climate change effects in high productivity regions.

Our indicator of climate change damages is derived from a drought based indicator, the Global Standardized Precipitation-Evapotranspiration Index (SPEI), which is a well established benchmark for capturing joint effects of precipitation, potential evaporation and temperature \citep{Begueria2014}. The SPEI index uses monthly precipitation and potential evapotranspiration data from the CRU weather database as an input variable. The SPEI data is available monthly as a globally gridded dataset, where each pixel has a half-degree resolution. Each monthly observation of the index informs on  deviations from the average available water (i.e. the presence and severity of droughts). Lower values of the index correspond to  larger deviations from average water availability conditions. Additionally, SPEI data also characterizes the persistence of drought, ranging from one to twelve months. In our analysis, we focus on persistent droughts affecting a substantial share of national agricultural good production. We rely on the index of droughts that are persistent over a time window of three months, and that are severe to exceptional (SPEI $< -1.5$), based on the classification of \cite{Leng2019}.

To quantify the impact of severe to exceptional droughts on agricultural production, we combine the SPEI data with spatially explicit data on agricultural production. The spatial production allocation model (SPAM) provides a well-established baseline of gridded agricultural production areas at a 5-arcminute resolution in the year 2000, obtained by harmonizing sub-national and national production statistics with satellite-based remote sensing data \citep{Balkovic2014}. SPAM provides growing area information on a wide range of crop types, with varying degrees of accuracy. Due to data limitations and to provide robustness for our assessment, we focus on the main crop types corn, rice, maize and wheat, which jointly cover 75 percent of the caloric content of global food production \citep{Roberts2010}. Observations on average yields in kilocalories per hectare under different levels of crop management intensification are obtained from the biophysical EPIC model \citep{VanderVelde2012}.\footnote{The EPIC model simulates potential crop growth under various biophysical and management conditions, which we transform by using FAO conversion values from tons per hectare for each of our main crops.} Combining data on crop production areas and yields provides us with a gridded dataset of potential production in kilocalories at a half-degree level.

Drought related events are assumed to only have an impact during growing season, which differs substantially by production regions. We use the AQUASTAT crop calendar to inform us on world-wide growing seasons. We aggregate the  percent of monthly kilocaloric crop production for eleven supra-national regions (see \cref{sec: app_regdat} for a definition of the climatic regions), which is subject to severe or exceptional droughts lasting for three or more months during growing season. 

\begin{figure}[t]
\begin{center}
\includegraphics[width=\textwidth]{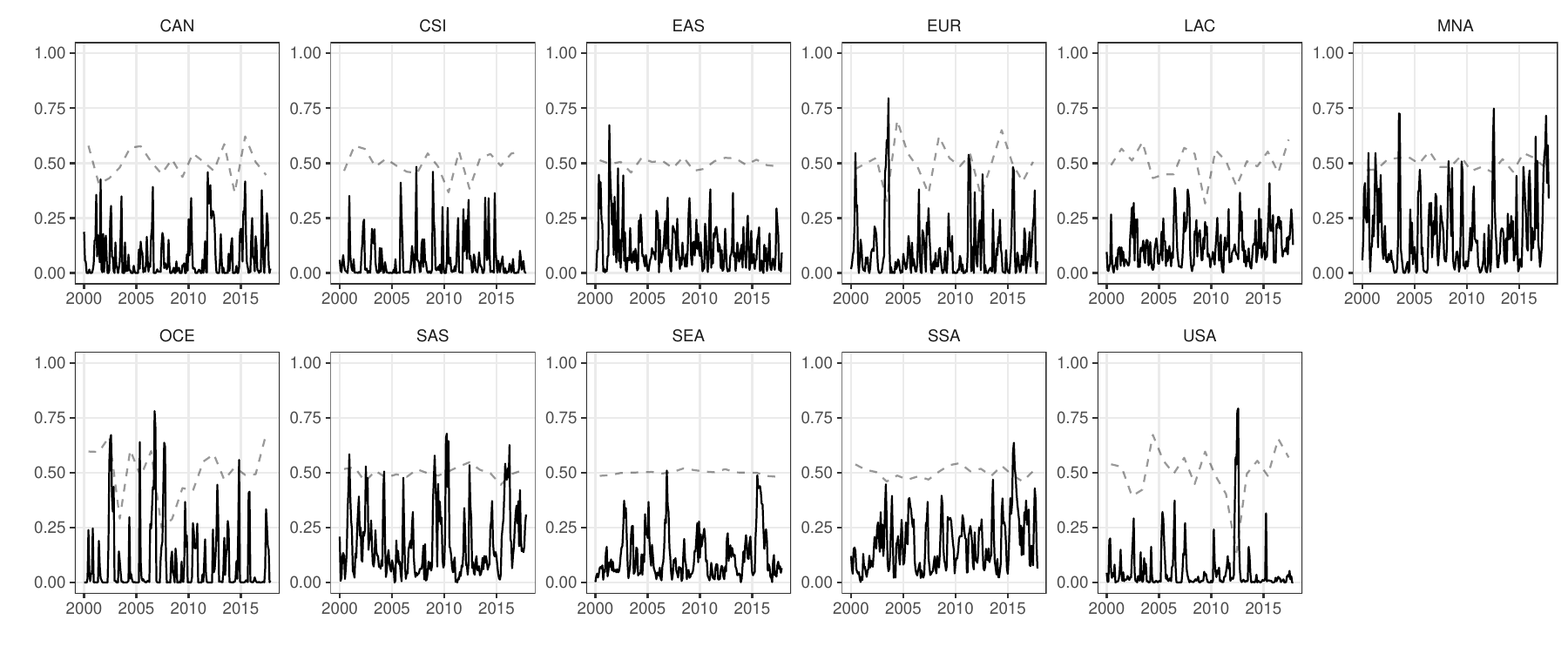}
\end{center}
\caption{Percent of caloric production under severe and exceptional three-month drought.}\vspace*{0.25em}
\footnotesize\textit{Notes}: Dashed lines correspond to detrended annual deviations from average kilocaric yield (Source: \textit{FAO}).
\label{fig:spei_yield}
\end{figure}

Figure \ref{fig:spei_yield} displays the resulting climate change index. The dashed lines correspond to annual demeaned and detrended yield in kilocalories per hectare for the four major crops (obtained from \textit{FAO}), centered at 0.5 for easier comparison. Inspection of regional results reveals that trends in annual yield shortfalls are captured well by the index. Particularly well-known severe drought events, such as the drought in the US in 2012/13 or in Australia (located in the OCE group) in 2007/08 are clearly visible. The index also picks up other climate-related stylized facts, such as the higher incidence of droughts and extreme climate events in the Middle East and North Africa (MNA), Sub-Saharan Africa (SSA) and South-East Asia (SAS).

\subsection{Model specification}
For the empirical study of this paper, we adapt the baseline model in \cref{sec: econometrics} along two dimensions. First, we include a block of $L$ quantities in a vector $\bm{y}_{0t}$, capturing dynamics on futures markets for energy and agricultural commodities. Here, we define $\bm{z}_t = (\bm{y}_{0t-1}',\hdots,\bm{y}_{0t-P}')'$ and $\bm{D}_i = (\bm{D}_{i1},\hdots,\bm{D}_{iP})$. Second, we introduce exogenous scalar time series $r_{jt}$ containing information on climate shocks for various aggregate regions indexed by $j=1,\hdots,\mathcal{J}$ to assess the contemporaneous responses of all endogenous variables in the system. Analogous to \autoref{eq:PVARstacked}, we use $\bm{x}_{it}= (\bm{y}'_{it-1},\dots, \bm{y}'_{it-P})', \bm{C}_i = (\bm{A}_{i1},\dots,\bm{A}_{iP}), \bm{x}_{-i, t}=(\bm{y}'_{-i, t-1}, \dots, \bm{y}'_{-i, t-P})' $ and $\bm{B}_i = (\bm{B}_{i1}, \dots, \bm{B}_{iP})'$. Note that we disregard the intercept term and deterministic trends for the sake of simplicity, however, including such terms is straightforward. The vector autoregressive process for countries $i=1,\hdots,N$ is given by
\begin{equation}
\bm{y}_{it} = \bm{C}_{i} \bm{x}_{it}+ \bm{B}_i \bm{x}_{-i, t} + \bm{D}_i\bm{z}_t + \sum_{j=1}^{\mathcal{J}}\bm{\gamma}_{ij} r_{jt} + \bm{\varepsilon}_{it}.\label{eq:appl_country}
\end{equation}
The climate shocks $r_{jt}$ enter the model as exogenous quantities, with the associated $K$-dimensional parameter vectors $\bm{\gamma}_{ij}$ capturing the contemporaneous response of all endogenous variables of country $i$ to a climate shock in region $j$. For the global quantities, we assume again a vector autoregressive process and also include information on the country-specific series. 

Specifically, the model for these variables reads
\begin{equation}
\bm{y}_{0t} = \bm{C}_0\bm{z}_t + \bm{B}_{01}\bm{y}_{t-1} + \hdots + \bm{B}_{0P}\bm{y}_{t-P} + \sum_{j=1}^{\mathcal{J}}\bm{\gamma}_{0j}r_{jt} + \bm{\varepsilon}_{0t},\label{eq:appl_global}
\end{equation}
with $\bm{C}_0 = (\bm{A}_{01},\dots,\bm{A}_{0P})$ and coefficient matrices $\bm{B}_{0p}$ associated with the $p$th lag of $\bm{y}_{t}$. Here, the $L$-dimensional vectors $\bm{\gamma}_{0j}$ again measure the contemporaneous effects of exogenous climate shocks on futures markets prices. 

In particular, the variables included in the vector of global futures market quantities $\bm{y}_{0t}$ and the country-specific macroeconomic time series in $\bm{y}_{it}$ are
\begin{align*}
\bm{y}_{0t} &= (\text{crudeoil},\text{gasoil},\text{corn},\text{rice},\text{soya},\text{soyoil},\text{wheat},\text{cotton},\text{ethanol},\text{hogs},\text{fcattle},\text{lcattle})',\\
\bm{y}_{it} &= (\text{CPI},\text{CPF},\text{ir},\text{prod},\text{XR2})'.
\end{align*}

The priors on the domestic coefficient matrices $\bm{C}_i$ in \autoref{eq:appl_country} are chosen as proposed for the general model and thus are assigned a sparse finite mixture distribution. For the dynamic interdependencies $\bm{B}_i$ and the parameters associated with futures prices $\bm{D}_i$, we choose the NG shrinkage prior set forth in \cref{sec: econometrics}. Here, we rely on country-specific global parameters, pooling information across variable types. For the futures market model given by \autoref{eq:appl_global}, we again achieve regularization by employing an analogous NG shrinkage prior. Let $\bm{b}_0 = \text{vec}(\bm{C}_0,\bm{B}_{01},\hdots,\bm{B}_{0P})$, with the $l$th element of $\bm{b}_0$ given by $b_{0l}$. Specifically, the prior is
\begin{equation*}
b_{0l}| \tau_{0l}, \xi_0 \sim \mathcal{N}\left(0, \frac{2 \tau^2_{0l}}{\xi_0}\right),\quad \tau^2_{0l} \sim \mathcal{G}(\vartheta_0, \vartheta_0), \quad \xi_i \sim \mathcal{G}(\mathfrak{c}_0, \mathfrak{c}_1),
\end{equation*}
with hyperparameters $\vartheta_0=0.1$, $\mathfrak{c}_0 = \mathfrak{c}_1 = 0.01$. For the coefficients measuring the contemporaneous effects of climate shocks on the endogenous variables of the system, we opt for a weakly informative prior setup using
\begin{equation*}
\bm{\gamma}_{0j} \sim \mathcal{N}(\bm{0},10\times\bm{I}_L),\quad \bm{\gamma}_{ij} \sim \mathcal{N}(\bm{0},10\times\bm{I}_K), \quad \text{for}~i=1,\hdots,N,~j=1,\hdots,\mathcal{J}.
\end{equation*}
All other priors are set in line with the general framework proposed in \cref{sec: econometrics}. It remains to specify the lag length $P$, that we set equal to two, and the number of factors. Here, we rely on the NG shrinkage prior setup involved in achieving regularization of the factor loadings matrix, which indicates that $q=8$ factors are sufficient. Choosing a different number of factors leaves the results qualitatively unchanged. We iterate the algorithm 30,000 times and discard the first 15,000 draws as burn-in. Posterior inference is obtained considering each third of the remaining 15,000 posterior draws.

\section{The impact of climate shocks}\label{sec: application}
\subsection*{Key features of the modeling approach}
\begin{figure}[ht]
\begin{center}
\includegraphics[width=\textwidth]{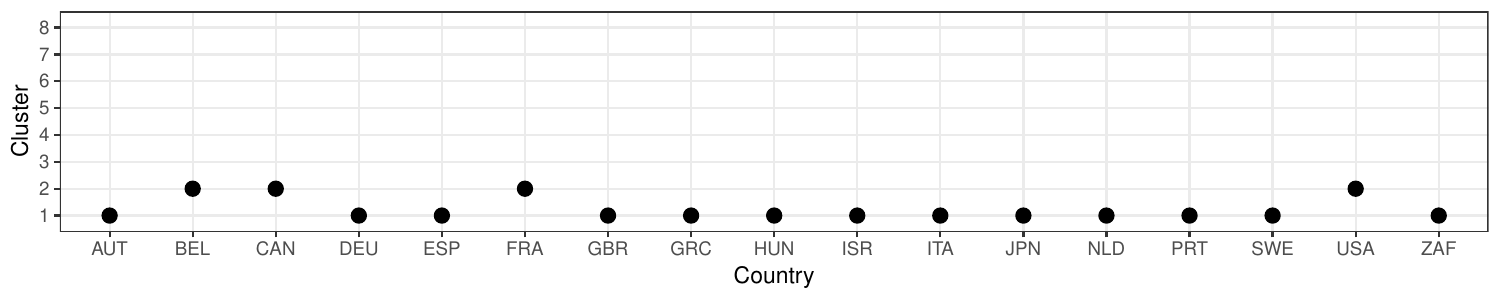}
\includegraphics[width=\textwidth]{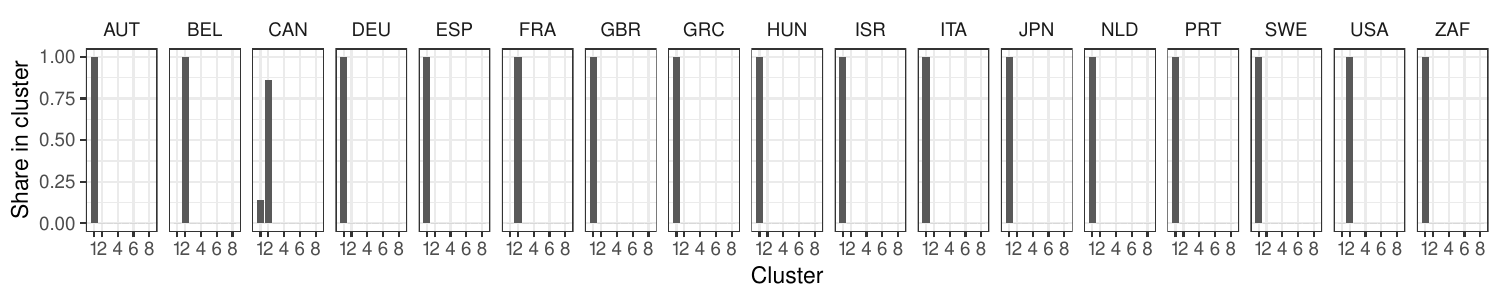}
\end{center}
\caption{Cluster allocation (top panel) and posterior inclusion probability per country (bottom panel).}
\label{fig:clusters}
\end{figure}

We start by investigating the cluster allocation of countries and the estimated number of clusters. Figure \ref{fig:clusters} displays the estimated regime allocation across countries in the upper panel. The estimated allocation indicates that within our sample of developed economies, the modeling approach yields two distinct clusters based on domestic dynamics governed by $\bm{c}_i$. In fact, computing the posterior distribution of the estimated number of regimes along the lines of \citet{malsiner2016model} yields a posterior probability for two clusters equal to unity. The smaller group contains four countries (Belgium, Canada, France and the United States), while the larger cluster features the remaining economies. In terms of inclusion probabilities (bottom panel), the cluster allocation of all countries is distinct. Most economies are included in their respective clusters in all iterations of the MCMC algorithm. An exception is Canada, which is included in the second group with an inclusion probability of 86 percent, and 14 percent for the first group. Assessing drivers of cluster allocation in terms of $\lambda_j$ and the posterior distribution of $\bm{\mu}_1-\bm{\mu}_2$ shows that main differences occur for the respective CPI equation.

\begin{figure}[ht]
\begin{center}
\includegraphics[width=\textwidth]{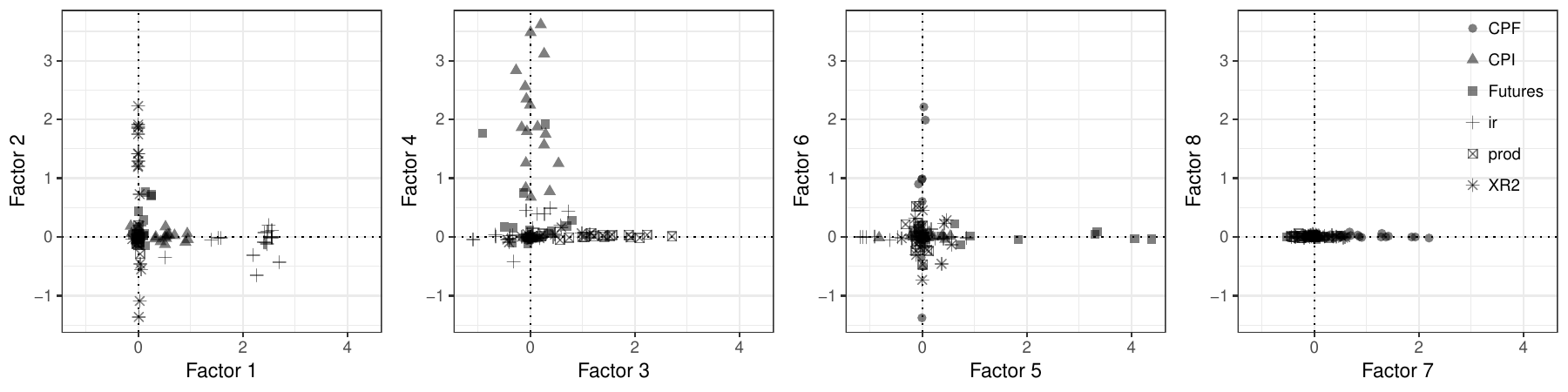}
\end{center}
\caption{Pairwise scatterplots of the median factor loadings.}\label{fig:facloads}
\end{figure}

In a second step, we assess the importance of SIs and thus comovements across variables types governed by the factor structure. Figure \ref{fig:facloads} shows pairwise scatterplots of the columns of the median estimates of the factor loadings matrix $\bm{L}$. This visualization serves to illustrate which series load most strongly on which factor, and to give a qualitative interpretation to the latent factors. The first factor appears to reflect comovements in interest rates, with most corresponding series exhibiting quantitatively large factor loadings. Analogously, the third factor tracks joint movements in industrial production across economies. Turning to factors shaping the error structure of commodities futures prices, we find the highest factor loadings for the second, fourth and fifth factor. Interestingly, the second factor appears to also drive the shocks for exchange rates, providing a link between foreign exchange rate and commodity markets. The fourth factor governs contemporaneous dynamics of commodities futures markets and consumer price inflation across economies, while the fifth factor shows the largest loadings for futures prices in terms of the respective magnitude of the loadings. Notice that the eighth factor only shows loadings close to zero, providing further evidence that this number is likely sufficient to capture dynamics in the variance-covariance structure of the model.

\begin{figure}[ht]
\begin{center}
\includegraphics[width=\textwidth]{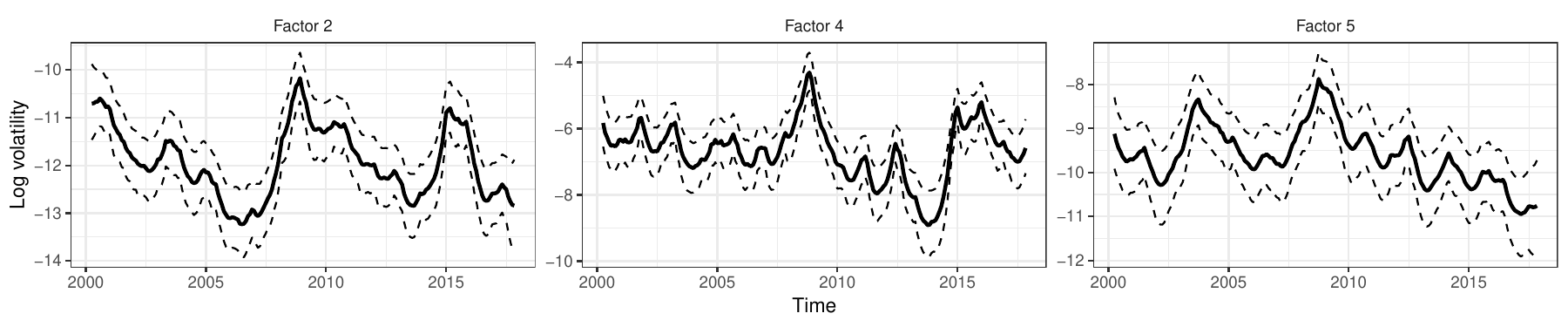}
\end{center}
\caption{Log volatilities of the factors with highest loadings for agricultural and commodities futures prices.}\vspace*{0.25em}
\footnotesize\textit{Notes}: The solid black line indicates the posterior median, while dashed lines refer to the 16th and 84th percentiles of the posterior distribution.
\label{fig:facvola}
\end{figure}

In the following discussion, we focus on three factors associated with the largest loadings on futures market prices. Figure \ref{fig:facvola} plots the full history of the log volatilities over the sample period. The volatilities with the highest loadings on commodity futures highlight multiple stylized facts. First, the peak of the volatilities in 2008 in all three plots coincides with the global financial crisis. However, note that fifth factor also reflects the so-called food price crisis, where food and energy commodities rose sharply, with associated increased volatility \citep{Headey2011}.  Additionally, the sharp increase in commodity futures volatilities in 2013 corresponds to severe droughts impacting the US. It should be pointed out that the magnitude of volatilities differs sharply across the factors, with the fourth factor -- most strongly associated with CPI and commodity futures -- exhibiting the largest values. Commodity futures are in fact seen as a possible hedge against inflation, providing an explanation for the shared volatility structure. 

\begin{figure}
\begin{center}
\includegraphics[width=\textwidth]{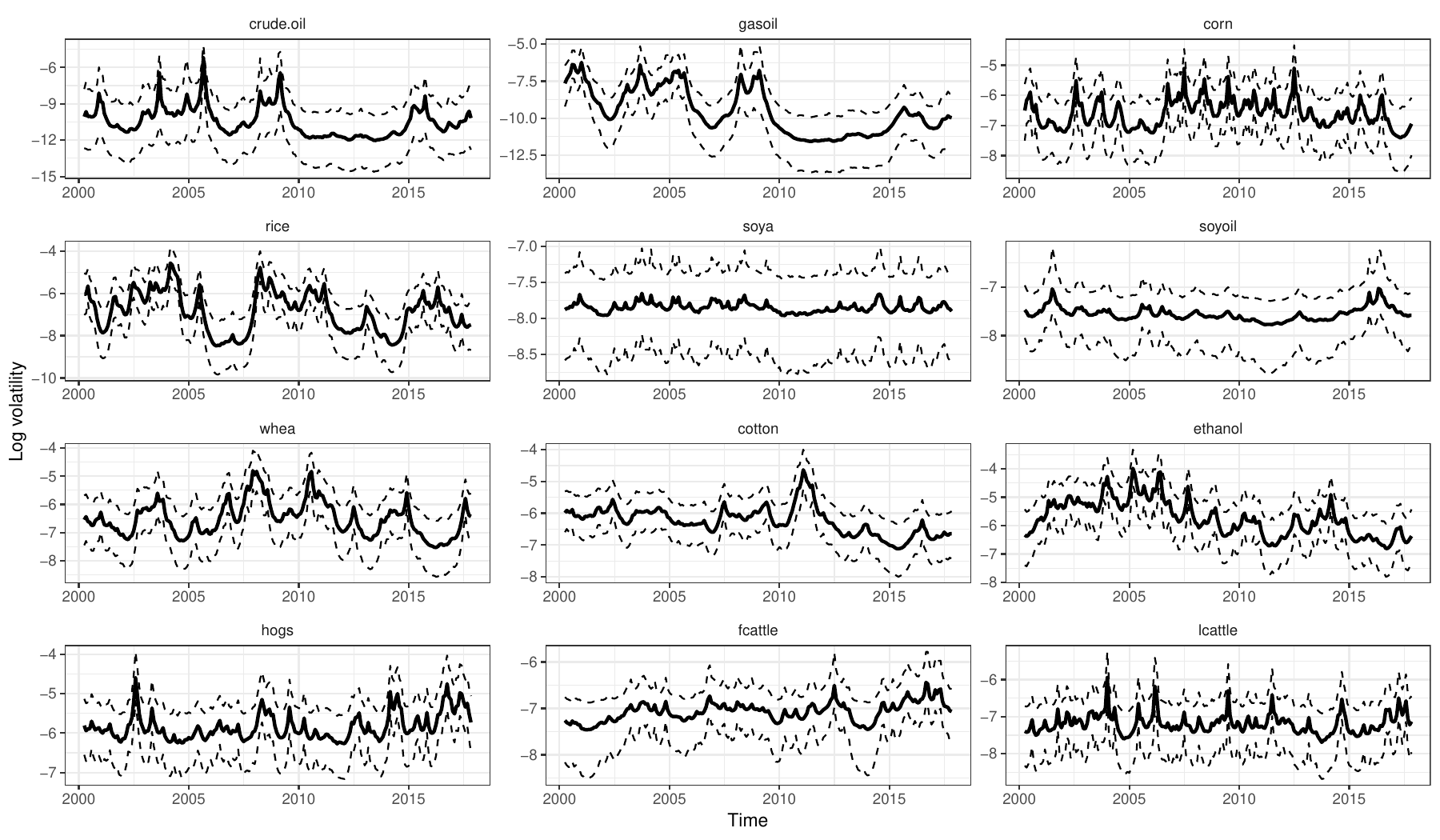}
\end{center}
\caption{Idiosyncratic log volatility components of commodities futures prices.}
\footnotesize\textit{Notes}: The solid black line indicates the posterior median, while dashed lines refer to the 16th and 84th percentiles of the posterior distribution.
\label{fig:idiovola}
\end{figure}

For the sake of completeness, we moreover present some evidence on series specific idiosyncrasies for commodities futures prices in \autoref{fig:idiovola}. A few points are worth noting here. First, we again find substantial differences in the magnitude of the volatilities across series. Second, evidence for time variation in the idiosyncratic components of the error terms is muted for soybeans and soyoil while we find substantial movements in the idiosyncratic log volatilities for the remaining series. Specifically, we do not find evidence for an increase in the idiosyncratic log volatilities of biofuels during the stock market and food price crisis of 2008/09. This implies that while energy and agricultural futures clearly share similar volatility patterns, the link is not necessarily provided by biofuel markets and related US policy \citep{Nazlioglu2011}. 

The idiosyncratic log volatilities of crude oil and gasoil exhibit similar patterns, especially after the global financial crisis. The 2015 volatility spike, which is largely due to supply side consolidation, especially of shale oil companies, can clearly be observed. The log volatilities of corn exhibit seasonal patterns. These represent a well-known fact of the US corn futures market and are connected to corn stock movements. Moreover, the 2013 drought evident in Fig. \ref{fig:spei_yield} is reflected as the highest volatility spike in corn futures. 

Closer inspection of idiosyncratic livestock log volatility patterns reveals that live cattle and hogs closely track the volatilities of crop markets as feeder cattle. This reflects the fact that both of these livestock futures relate to animals which have to reach full maturity and are tightly linked to global feed prices. Note, that some pronounced volatility spikes in corn, cotton, and  ethanol futures are tracked with a slight delay by the hogs and live cattle markets, due to these crops being a major food source for hogs and cattles. The increase in the log volatilities of the livestock market since 2016 can be clearly observed. Initially, changing CME future market regulations were blamed for this increase, however, the trend has persisted since then despite regulatory countermeasures, and the causes are subject of ongoing research.

\subsection*{Climate change impact on global future prices}
In the following, we consider the impact of climate-related drought shocks on financial markets. The empirical literature emphasizes spillovers from energy to agricultural markets as causal effects for rising food prices \citep{Lucotte2016,Nazlioglu2011,Baumeister2013}. The dynamic responses of a wide range of commodity futures to regionally located climate shocks provide evidence on the prevalence of strong linkages between global commodity prices and climate change.

\begin{figure}[t]
\begin{center}
\includegraphics[width=0.49\textwidth]{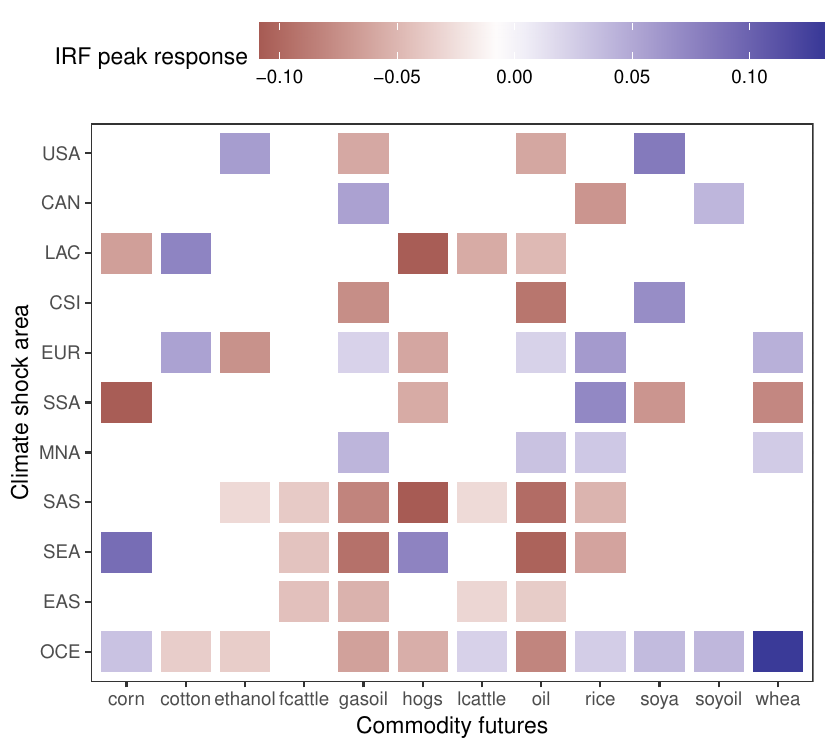}
\includegraphics[width=0.49\textwidth]{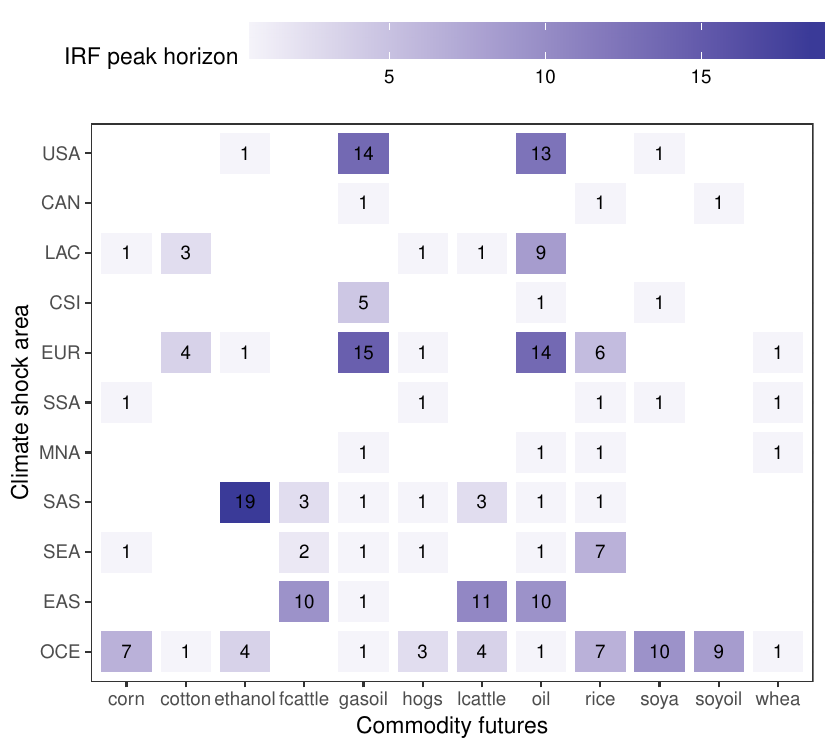}
\end{center}
\caption{Peak impulse response functions (left panel) and timing of the peak response in months after impact (right panel) for commodities futures prices.}\vspace*{0.25em}
\footnotesize\textit{Notes}: The shading indicates the respective magnitude and timing of the response. Insignificant impulse responses, based on the 16th and 84th percentiles of the posterior distribution covering zero, are left blank.
\label{fig:CME_shocks}
\end{figure}

Figure \ref{fig:CME_shocks} presents the peak endogenous responses to exogenous climate change shocks (left panel), measured in percent of potential kilocaloric production under severe drought. This is displayed alongside the timing of the peak response in months (right panel). Insignificant impulse responses, i.e. responses where the 16th and 84th percentiles of the posterior distribution cover zero, are displayed as blank.

 A first inspection of the graph reveals that droughts can induce both positive and negative reactions in the futures markets. Positive price shocks are a direct result of tightening agricultural supply. Negative price shocks can stem from various causes. First and foremost, negative price movements result from droughts causing a decrease in demand. The literature suggests that these effects play a central role in lower-income regions, such as Sub-Saharan Africa, where wide-spread subsistence level farming coupled with climate change leads to lower demand for imported agricultural goods \citep{Minot2014,Amare2018}. A further prevalent explanation for a short-term price decrease is the overshooting behavior of agricultural commodity prices, represented in theoretical models by \cite{Dornbusch1976} and \cite{Frankel1986}. Finally, a decrease in prices as a result of a localized drought shock might be caused by production shifting to other major producing regions not affected by drought. This is evident in the case of rice, where 92 percent of global production is localized equally in South-East Asia, Southern Asia and Eastern Asia.

In the case of a drought shock in the US, our results suggest that future prices of corn, wheat, and rice are not affected. This is likely due to the US possessing significant stocks of these commodities. Indeed, major future price shocks coupled with drought events in the US happened only in 2012, when stock to use ratios were low. Our model, however, seems to provide evidence for the increased interrelationship between biofuel and energy markets. This is evidenced by a drought shock leading to a sharp, short term spike in prices of soybeans and ethanol. Likely due to the overshooting behavior of commodity markets, this translates into a dip in crude oil and gasoil future prices peaking after  13 and 14 months, respectively. This supports evidence from the literature arguing for the existence of increased dependence of energy and agricultural markets \citep{Nazlioglu2011,Lucotte2016}. The peak impulse response functions for Canada (CAN) further support this, as a drought related shock in Canada in turn increases soyoil prices and also leads to a simultaneous peak in gasoil prices, due to the country being a major global biofuel producer. 

A drought related shock to agricultural production areas in Latin American countries (LAC) leads to an immediate dip in corn, live cattle, and hogs future prices. This is likely due to a quick sell-off of live cattle and hogs, due to the projected increase in prices of feed (mainly corn), which also results in a decrease of feed demand. Additionally, climate related shocks -- especially in South and Central America -- are thought to lead to a conversion of cropland areas into pastures \citep{Headey2011}, which in turn increases global supply of livestock products. The concurrent increase in global cotton prices which peaks after three months provides further evidence for this, as cotton waste (a major by product of cotton production) is a predominant feed for livestock.

The peak impulse responses to a drought shock in Europe (EUR) suggests an increase in rice, wheat and cotton prices, coupled with an increase in crude oil and gasoil prices peaking after a 14 month period. This, together with an immediate decrease in ethanol and hogs futures, suggests that faced with drought farmers switch production to ethanol and livestock production, which are both not as sensitive to weather related events. The resulting increase in biodiesel production leads to a slight decrease in global oil prices due to over-supply, which further underlines the evidence for cointegration of energy and agricultural markets \citep{Nazlioglu2011}. 

Former USSR countries have been a major producer and consumer of soybeans prior to the collapse of large-scale production in the late 1990s. As a result, Russia and Western Asian (CSI) countries are major importers of soybeans, and have started expanding production only in the last decade \citep{Headey2011}. Therefore, a drought shock translates to a sharp increase in global soy markets, as evidenced in our results. The significant and persistent decrease in global oil future prices is likely due to droughts being coupled with higher temperatures, which can significantly lower heating costs in cold climate regions.

Turning our attention to the results of drought shocks in the Middle-East and North Africa (MNA), we provide evidence of sharp increases in wheat, rice, crude oil, and gasoil prices. This is related to the drought impacting domestic production. Therefore, the shortfall in consumption results in an increase of global demand for rice and wheat. The increase in crude oil and gasoil prices is due to higher transportation costs and increases in the cost of oil production itself. For shocks in Sub-Saharan Africa (SSA), our results suggest a decrease in corn, hogs, soybean and wheat prices, coupled with an increase in the global futures price for rice. Household level surveys such as \cite{Minot2014} and \cite{Amare2018} provide evidence for a drop in demand for multiple imported goods due to a decrease in household income, which is supported by our results. While SSA accounts for only two percent of global rice production, they are a major per capita consumer of rice products, and the majority of local production is also consumed locally \citep{Headey2011}. Thus a drought shock turns into higher imports and, in turn, an increase in global prices.

Considering the results for droughts in Southern Asia (SAS), we find decreases in prices across a wide range of futures. Since the region itself encompasses developing economies with a low per capita income but a large share of global production, this provides further support of a change in demand in relatively poorer countries as a result of climate shocks \citep{Minot2014,Amare2018}. The persistent dip in oil prices, petering out after a period of approximately two years is evidence of the major role of transportation costs: a shortfall of exports leads to less demand for fuel \citep{Headey2011}. South-East Asia (SEA) presents an interesting case, as it experienced major expansions of palm oil plantations -- used both in fuel production and as biofuels -- to the detriment of cropland and pastures in the recent decade. This is reflected in our results by a sharp increase in corn and livestock prices, coupled with decreasing crude and gasoil future prices. Finally, our results for the major producing region of Eastern Asia (EAS), encompassing China, Japan, and South Korea, underline overshooting behavior of commodity prices, where the initial drought shock leads to an overproduction of feed, and thus to a lowering of livestock prices peaking after ten months, which corresponds to a full agricultural growing cycle in the region.

Drought shocks in Oceania (OCE), covering the high output economies New Zealand and Australia, yield similar responses as in Europe. The shortfall in domestic production leads to higher demand and a sharp increase in global wheat, corn, soy oil, and rice prices. These price spikes are persistent and in the case of soya, rice and corn, peak only after seven months. Due to the increase in feed prices live cattle prices also increase, peaking after four months. Feeder cattle is not significantly impacted and lean hogs future prices decrease, likely due to producers selling early to avoid increased production costs. The decrease in global oil prices points to a decrease in exports.

\subsection*{Impacts on high-income economies}
\begin{figure}[hp]
\begin{center}
\includegraphics[width=0.82\textwidth]{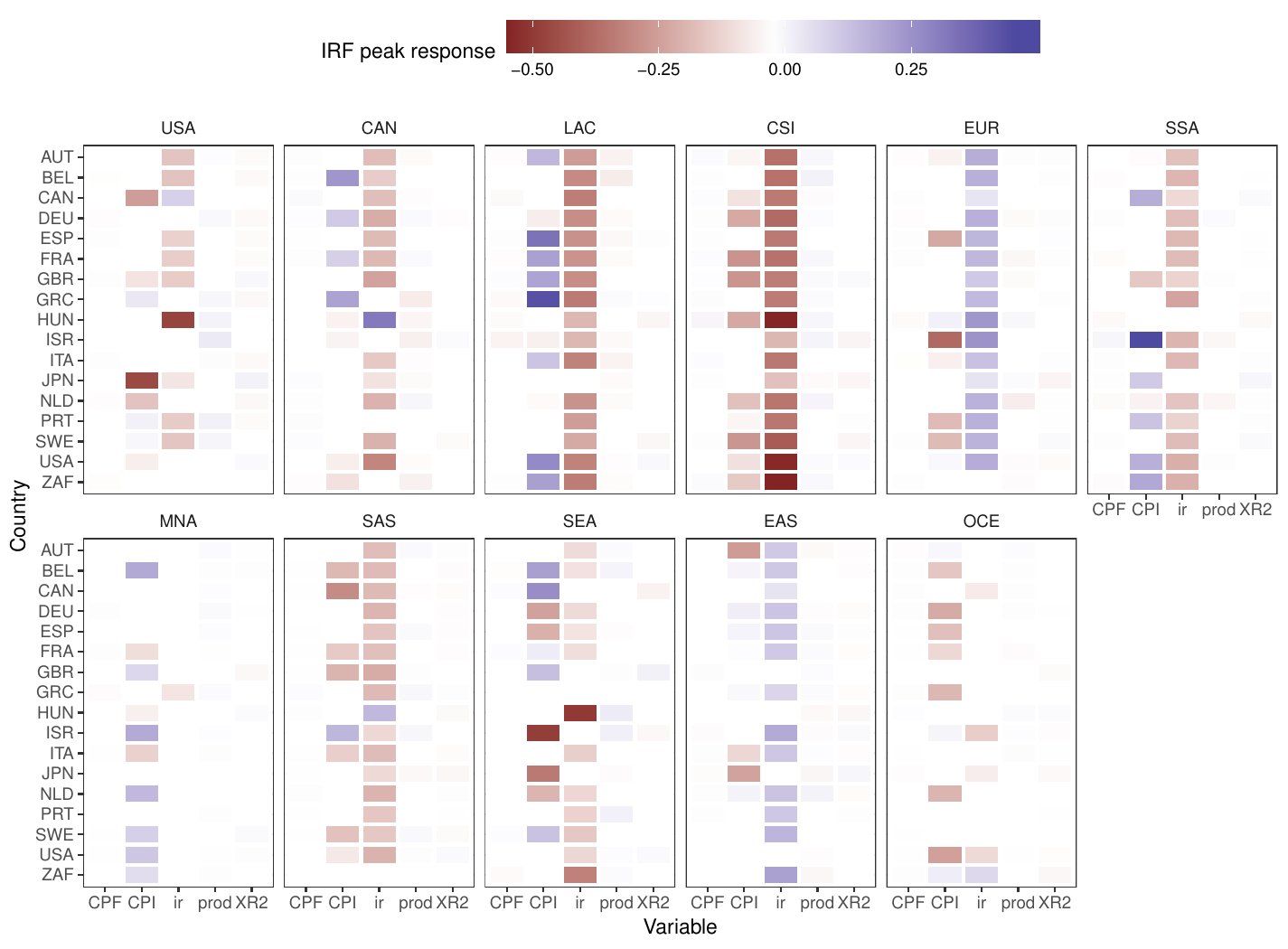}
\includegraphics[width=0.82\textwidth]{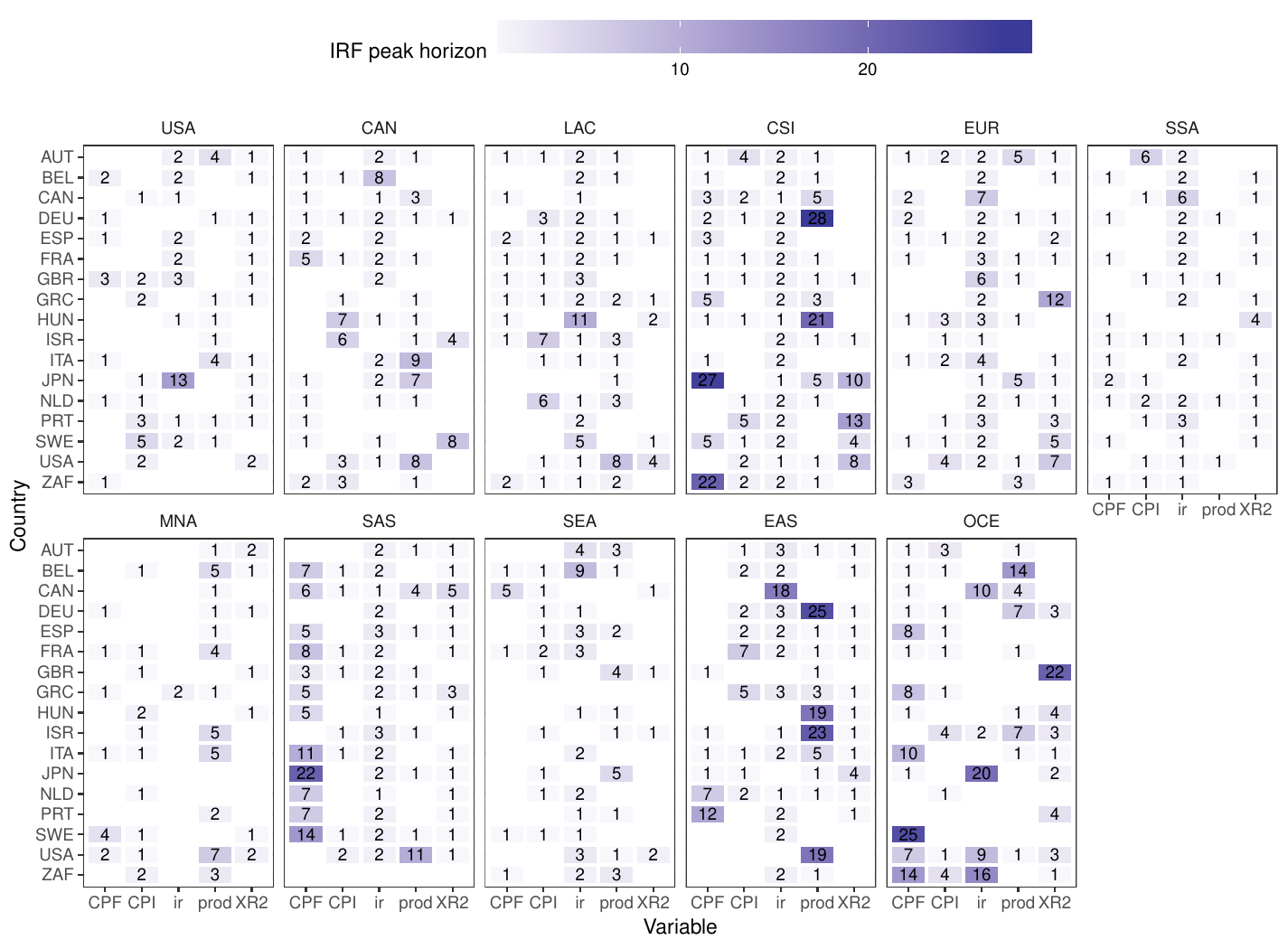}
\end{center}
\caption{Peak impulse response functions (upper panel) and timing of the peak response in months after impact (lower panel) across countries and variable types.}\vspace*{0.25em}
\footnotesize\textit{Notes}: The shading indicates the respective magnitude and timing of the response. Insignificant impulse responses, based on the 16th and 84th percentiles of the posterior distribution covering zero, are left blank.
\label{fig:country_peak_shocks}
\end{figure}

In order to obtain a more detailed picture of how global drought shocks play out in developing economies, we now consider the country-specific results of our model. Figure \ref{fig:country_peak_shocks} displays  peak  responses to observed, exogenous climate change shocks (upper panel). Each subpanel of the figure contains peak responses for all drought shock in the eleven global climatic regions. Insignificant impulse responses (under the 16th and 84th percentile) are again displayed as blank. Analogously, the bottom panel displays the timing of the peak responses in months. 

A first inspection of the figure reveals that droughts indeed exhibit spillovers on price changes, interest rates, and economic output. The magnitude, direction, and significance of these shocks is dependent on a countries' overall dependence on agricultural goods, as well as geographic proximity and trade ties to the drought impacted climatic region. Nonetheless, some regularities can be readily observed. Exchange and interest rate responses of Euro area countries exhibit comovements, due to common monetary policy. Additionally, the responses of interest rates display significant comovement across the sample. The direction of this comovement seems to strongly depend on which climatic region is impacted by drought. This can be explained by the fact that the countries of interest represent developed economies with homogeneous choices of trading partners amongst the eleven climatic regions \citep{Baker2018}. While overall food prices exhibit little movement, the responses of short term interest rates can be seen as a compensation of shocks that would increase food prices. This is supported by recent literature documenting the interrelationship of food CPI and interest rates \citep{Akram2009,DeNicola2016}. Finally, it is worth mentioning that the peak responses manifest within the first 12 months in countries that were directly impacted by the drought. Spillover to other countries typically materialize later on. 

As a response to an exogenous drought shock in the US, we observe a minor  increase of the real effective exchange rate of the US dollar, which is consistent with the US being a large importer of food stocks. This indicates that the shortfall of production is matched by imports, coupled with a decrease in prices. Interestingly, the exchange rate of Canada -- the largest trading partner of the US -- does not change significantly, reflecting that agricultural commodities are imported from elsewhere. However, Canada experiences a short decrease in inflation, which is matched by a decrease in interest rates consistent with expansionary monetary policy counteracting disinflationary pressures. The real effective exchange rates of almost all Euro area economies and Israel exhibit a slight dip, coupled with decreases in short-term interest rates and a one to five month spike in output. This indicates a direct increase in production to cover the US shortfalls. The responses in inflation are mixed, pointing to a slight short-term increase in Sweden, Portugal and Greece. Food prices in relation to overall prices decrease in most European countries. Israel exhibits a sharp spike in inflation growth and a sharp increase in food prices, lasting for approximately five months. 

Focusing on the domestic response to a drought shock in Canada (CAN), we observe that both the inflation rate and food prices experience a short spike, where food prices exceed average prices by over one percent initially. This is matched by a decrease in interest rates, and eventually a statistically significant decline in production. Canada's largest trading partner, the US, exhibits similar patterns, albeit without increases in relative food prices and with a more pronounced decrease in production. The timing of the declines in production is concurrent with the increase in energy and biodiesel futures. Price and output fluctuations can be observed throughout the European economies around the first two months, reflecting adjustments to the economy to a drought shock in a large ethanol producing country. Note that major importers of Canadian products, such as France, also experience spike in food prices between one to six months after the climate shock hits the system.

The peak impulse responses to a climate change shock in Latin America (LAC) indicate a negative impact on interest rates across almost all countries of the sample, accompanied by a short period of declines in output. This reflects the large contribution of Latin American countries to crop and livestock production globally \citep{Headey2011}. Moreover, this coincides with the observed rise in cotton prices and the implied shortfalls of livestock feed, but also with the dip in livestock prices. Major trading partners like the US and some European countries increase imports. Additionally, across the Euro area food prices relative to overall price levels dip significantly after the shock, with the exception of Britain, Spain and South Africa, all of whom are major importers of feed products.

The real effective exchange rate and short-term interest rate responses to a drought shock in Europe (EUR) provide evidence that all European economies (with the exception of the United Kingdom and Hungary) significantly increase imports to match the shortfalls in agricultural production. The responses in output within Europe are mixed, with initial fluctuations in both directions. Small, but significant dips in output beyond the ten-month horizon can be observed in Austria, Germany and Hungary. Note that food prices decrease across all countries, a result that coupled with the response in short-term interest rates supports evidence by \cite{Akram2009}, who identifies strong linkages between food price and interest rates. The responses in the US to a European drought shock demonstrate an increase in exports, coupled with an initial drop in production. This phenomenon occurs at the same time as global future prices for wheat and rice increase, pointing to shortages in food availability as a possible explanation.

The impacts of a drought shock in Russia and Western Asia (CSI) are characterized by a reduction of short-term interest rates, combined with short-term food price and output spikes in major trading partners. Note that positive impacts on production last up to two years in most European countries. This is closely related to our findings that drought shocks in the CSI region induce a long-term dip in crude oil and gasoil prices \citep{Lucotte2016}.

Consistent with the  impulse response functions of global commodity futures, the results from a drought shock in Sub-Saharan Africa (SSA) correspond to a decrease in demand in the affected region. Therefore, almost all countries in our sample exhibit a statistically significant increase in real effective exchange rates, which is consistent with an increase in imports. This is accompanied in most countries by a dip in food prices, except Japan and Israel, both of which have substantial agricultural import ties with Sub-Saharan Africa. South Africa is not significantly affected in terms of output or exchange rate fluctuations. This underlines findings of \cite{Amare2018} who argue that demand shifts due to droughts are only in effect in lower-income countries as opposed to developed economies. Droughts in Middle-East and Northern Africa (MNA)  cause a significant, albeit small, reaction on output and inflation in almost all countries under consideration. 

Examining drought shocks in Eastern Asia (EAS), we observe that Japan faces a shortfall in production and a negative reaction of output. Moreover, we find a positive exchange rate response, which corresponds to an increase in imports. Most countries in turn -- with the exception of Israel and Japan -- exhibit significantly lower real effective exchange rates, that is, an increase in exports. The results highlight that countries such as the United States, Austria, Germany, Hungary, and South Africa experience a small delayed (after a horizon of twelve months), albeit significant decrease in industrial production. This might be an indication of the overshooting behavior of commodity markets, since many of these countries expand their production in response to the drought shock in Asia.

\section{Concluding remarks}
In this paper, we focus on the highly policy relevant question of assessing the impacts and transmission channels of climate change shocks on a set of high-income OECD economies. For this purpose we measure climate change related shocks using a novel index measuring the percentage of agricultural production under severe, persistent drought, constructed by using spatially explicit datasets. Moreover, we specifically focus on macroeconomic quantities such as output, interest and exchange rates, as well as food prices and inflation. Moreover, our approach controls for the crucial role of global commodity markets in the agricultural sector. In order to efficiently cope with the large number of variables, we develop a PVAR model that pools information across countries using a sparse finite mixture of Gaussians prior on the domestic, country-specific coefficients. We control for the existence of dynamic interdependencies by relying on a global-local shrinkage prior to stochastically select non-zero relationships across countries and variable types. Static interdependencies are parsimoniously modeled through a factor stochastic volatility specification of the error variance-covariance matrix.

Our results demonstrate the impact of climate change, both on global commodity markets, as well as among country-specific macroeconomic indicators. The main impacts of climate shocks are reflected in interest rate and inflation movements, in line with US-based studies by \cite{Akram2009} and \cite{Cashin2017a}. This indicates that monetary policy could play a central role in minimizing climate change related spillovers. Additionally, we provide evidence for climate change impacts on output of high-income countries, even if the associated drought events manifest in a different part of the world. The global commodity market results corroborate the findings of \cite{Nazlioglu2012} and \cite{Lucotte2016}, who also find strong evidence for the interdependence between energy and agricultural markets coupled with an increased global demand for biofuels.

\small{\setstretch{0.85}
\addcontentsline{toc}{section}{References}
\bibliographystyle{bibtex/custom.bst}
\bibliography{bibtex/favar,bibtex/mpShocks,bibtex/Mendeley}}\normalsize

\begin{thebibliography}{73}
\newcommand{\enquote}[1]{``#1''}
\providecommand{\natexlab}[1]{#1}

\bibitem[{Aguilar and West(2000)}]{aguilar2000bayesian}
\textsc{Aguilar O, and West M} (2000), \enquote{Bayesian dynamic factor models
  and portfolio allocation,} \emph{Journal of Business \& Economic Statistics}
  \textbf{18}(3), 338--357.

\bibitem[{Akram(2009)}]{Akram2009}
\textsc{Akram QF} (2009), \enquote{{Commodity prices, interest rates and the
  dollar},} \emph{Energy Economics} \textbf{31}(6), 838--851.

\bibitem[{Allenby \emph{et~al.}(1998)Allenby, Arora, and
  Ginter}]{allenby1998heterogeneity}
\textsc{Allenby GM, Arora N, and Ginter JL} (1998), \enquote{On the
  heterogeneity of demand,} \emph{Journal of Marketing Research}
  \textbf{35}(3), 384--389.

\bibitem[{Amare \emph{et~al.}(2018)Amare, Jensen, Shiferaw, and
  Ciss{\'{e}}}]{Amare2018}
\textsc{Amare M, Jensen ND, Shiferaw B, and Ciss{\'{e}} JD} (2018),
  \enquote{{Rainfall shocks and agricultural productivity: Implication for
  rural household consumption},} \emph{Agricultural Systems}
  \textbf{166}(June), 79--89.

\bibitem[{Baffes and Haniotis(2010)}]{Baffes2010}
\textsc{Baffes J, and Haniotis T} (2010), \enquote{{Placing the recent
  commodity boom into perspective},} in \enquote{Food prices and rural
  poverty,} 40--70, Washington DC: World Bank.

\bibitem[{Baker \emph{et~al.}(2018)Baker, Havlik, Beach, Lecl{\`{e}}re, Schmid,
  Valin, Cole, Creason, Ohrel, and McFarland}]{Baker2018}
\textsc{Baker J, Havlik P, Beach R, Lecl{\`{e}}re D, Schmid E, Valin H, Cole J,
  Creason J, Ohrel S, and McFarland J} (2018), \enquote{{Evaluating the effects
  of climate change on US agricultural systems: sensitivity to regional impact
  and trade expansion scenarios},} \emph{Environmental Research Letters}
  \textbf{13}, 1--48.

\bibitem[{Balkovi{\v{c}} \emph{et~al.}(2014)Balkovi{\v{c}}, van~der Velde,
  Skalsk{\'{y}}, Xiong, Folberth, Khabarov, Smirnov, Mueller, and
  Obersteiner}]{Balkovic2014}
\textsc{Balkovi{\v{c}} J, van~der Velde M, Skalsk{\'{y}} R, Xiong W, Folberth
  C, Khabarov N, Smirnov A, Mueller ND, and Obersteiner M} (2014),
  \enquote{{Global wheat production potentials and management flexibility under
  the representative concentration pathways},} \emph{Global and Planetary
  Change} \textbf{122}, 107--121.

\bibitem[{Baumeister and Peersman(2013)}]{Baumeister2013}
\textsc{Baumeister C, and Peersman G} (2013), \enquote{{The role of
  time-varying price elasticities in accounting for volatility changes in the
  crude oil market},} \emph{Journal of Applied Econometrics} \textbf{28}(7),
  1087--1109.

\bibitem[{Beguer{\'{i}}a \emph{et~al.}(2014)Beguer{\'{i}}a, Vicente-Serrano,
  Reig, and Latorre}]{Begueria2014}
\textsc{Beguer{\'{i}}a S, Vicente-Serrano SM, Reig F, and Latorre B} (2014),
  \enquote{{Standardized Precipitation Evapotranspiration Index (SPEI)
  revisited: Parameter fitting, evapotranspiration models, tools, datasets and
  drought monitoring},} \emph{International Journal of Climatology}
  \textbf{34}, 3001--3023.

\bibitem[{Bhattacharya \emph{et~al.}(2015)Bhattacharya, Pati, Pillai, and
  Dunson}]{bhattacharya2015dirichlet}
\textsc{Bhattacharya A, Pati D, Pillai NS, and Dunson DB} (2015),
  \enquote{Dirichlet--Laplace priors for optimal shrinkage,} \emph{Journal of
  the American Statistical Association} \textbf{110}(512), 1479--1490.

\bibitem[{Burke and Tanutama(2019)}]{Burke2019}
\textsc{Burke M, and Tanutama V} (2019), \enquote{{Climatic constraints on
  aggregate economic output},} \emph{NBER Working Paper} \textbf{25779}.

\bibitem[{Canova and Ciccarelli(2004)}]{canova2004forecasting}
\textsc{Canova F, and Ciccarelli M} (2004), \enquote{Forecasting and turning
  point predictions in a Bayesian panel VAR model,} \emph{Journal of
  Econometrics} \textbf{120}(2), 327--359.

\bibitem[{Canova and Ciccarelli(2009)}]{canova2009estimating}
---{}---{}--- (2009), \enquote{Estimating multicountry VAR models,}
  \emph{International Economic Review} \textbf{50}(3), 929--959.

\bibitem[{Canova and Ciccarelli(2013)}]{canova2013panel}
---{}---{}--- (2013), \enquote{Panel Vector Autoregressive Models: A Survey,}
  in \enquote{VAR Models in Macroeconomics--New Developments and Applications:
  Essays in Honor of Christopher A. Sims,} 205--246, Emerald Group Publishing
  Limited.

\bibitem[{Cashin \emph{et~al.}(2017)Cashin, Mohaddes, and Raissi}]{Cashin2017a}
\textsc{Cashin P, Mohaddes K, and Raissi M} (2017), \enquote{{Fair weather or
  foul? The macroeconomic effects of El Ni{\~{n}}o},} \emph{Journal of
  International Economics} \textbf{106}, 37--54.

\bibitem[{Crespo~Cuaresma \emph{et~al.}(2016)Crespo~Cuaresma, Feldkircher, and
  Huber}]{cuaresma2016forecasting}
\textsc{Crespo~Cuaresma J, Feldkircher M, and Huber F} (2016),
  \enquote{Forecasting with global vector autoregressive models: A Bayesian
  approach,} \emph{Journal of Applied Econometrics} \textbf{31}(7), 1371--1391.

\bibitem[{de~Nicola \emph{et~al.}(2016)de~Nicola, {De Pace}, and
  Hernandez}]{DeNicola2016}
\textsc{de~Nicola F, {De Pace} P, and Hernandez MA} (2016),
  \enquote{{Co-movement of major energy, agricultural, and food commodity price
  returns: A time-series assessment},} \emph{Energy Economics} \textbf{57},
  28--41.

\bibitem[{Dees \emph{et~al.}(2007)Dees, Di~Mauro, Pesaran, and
  Smith}]{Dees2007a}
\textsc{Dees S, Di~Mauro F, Pesaran HM, and Smith LV} (2007),
  \enquote{{Exploring the international linkages of the euro area: A global VAR
  analysis},} \emph{Journal of Applied Econometrics} \textbf{22}(1), 1--38.

\bibitem[{Doan \emph{et~al.}(1984)Doan, Litterman, and Sims}]{Doan1984}
\textsc{Doan TR, Litterman BR, and Sims CA} (1984), \enquote{{Forecasting and
  conditional projection using realistic prior distributions},}
  \emph{Econometric Reviews} \textbf{3}(1), 1--100.

\bibitem[{Dornbusch(1976)}]{Dornbusch1976}
\textsc{Dornbusch R} (1976), \enquote{{Expectations and exchange rate
  dynamics},} \emph{Journal of Political Economy} \textbf{84}(6), 1161--1176.

\bibitem[{Enders and Holt(2014)}]{Enders2014}
\textsc{Enders W, and Holt MT} (2014), \emph{{The Evolving Relationships
  between Agricultural and Energy Commodity Prices: A Shifting-Mean Vector
  Autoregressive Analysis}}, University of Chicago Press.

\bibitem[{FAO(2017)}]{FAO2017}
\textsc{FAO} (2017), \emph{{The Future of Food and Agriculture: Trends and
  Challenges}}, Rome: FAO.

\bibitem[{Feldkircher and Huber(2016)}]{feldkircher2016international}
\textsc{Feldkircher M, and Huber F} (2016), \enquote{The international
  transmission of US shocks?evidence from Bayesian global vector
  autoregressions,} \emph{European Economic Review} \textbf{81}, 167--188.

\bibitem[{Frankel(1986)}]{Frankel1986}
\textsc{Frankel JA} (1986), \enquote{{Expectations and commodity price
  dynamics: The overshooting model},} \emph{American Journal of Agricultural
  Economics} \textbf{68}(2), 344--348.

\bibitem[{Fr{\"u}hwirth-Schnatter(2001)}]{fruhwirth2001markov}
\textsc{Fr{\"u}hwirth-Schnatter S} (2001), \enquote{Markov chain Monte Carlo
  estimation of classical and dynamic switching and mixture models,}
  \emph{Journal of the American Statistical Association} \textbf{96}(453),
  194--209.

\bibitem[{Fr{\"u}hwirth-Schnatter(2006)}]{fruhwirth2006finite}
---{}---{}--- (2006), \emph{Finite mixture and Markov switching models},
  Springer Science \& Business Media.

\bibitem[{Fr{\"u}hwirth-Schnatter and Kaufmann(2008)}]{frohwirth2008model}
\textsc{Fr{\"u}hwirth-Schnatter S, and Kaufmann S} (2008), \enquote{Model-based
  clustering of multiple time series,} \emph{Journal of Business \& Economic
  Statistics} \textbf{26}(1), 78--89.

\bibitem[{Fr{\"u}hwirth-Schnatter \emph{et~al.}(2004)Fr{\"u}hwirth-Schnatter,
  T{\"u}chler, and Otter}]{fruhwirth2004bayesian}
\textsc{Fr{\"u}hwirth-Schnatter S, T{\"u}chler R, and Otter T} (2004),
  \enquote{Bayesian analysis of the heterogeneity model,} \emph{Journal of
  Business \& Economic Statistics} \textbf{22}(1), 2--15.

\bibitem[{Frühwirth-Schnatter(2011)}]{fruhwirth2011label}
\textsc{Frühwirth-Schnatter S} (2011), \enquote{Dealing with Label Switching
  under Model Uncertainty,} in \enquote{Mixtures: Estimation and Applications,}
  213--239, John Wiley \& Sons, Ltd.

\bibitem[{Garcia \emph{et~al.}(2015)Garcia, Irwin, and Smith}]{Garcia2015}
\textsc{Garcia P, Irwin SH, and Smith A} (2015), \enquote{{Futures Market
  Failure?}} \emph{American Journal of Agricultural Economics} \textbf{97}(1),
  40--64.

\bibitem[{Gaupp \emph{et~al.}(2017)Gaupp, Pflug, Hochrainer-Stigler, Hall, and
  Dadson}]{Gaupp2017}
\textsc{Gaupp F, Pflug G, Hochrainer-Stigler S, Hall J, and Dadson S} (2017),
  \enquote{{Dependency of Crop Production between Global Breadbaskets: A Copula
  Approach for the Assessment of Global and Regional Risk Pools},} \emph{Risk
  Analysis} \textbf{37}(11), 2212--2228.

\bibitem[{Georgiadis(2015)}]{georgiadis2015examining}
\textsc{Georgiadis G} (2015), \enquote{Examining asymmetries in the
  transmission of monetary policy in the euro area: Evidence from a mixed
  cross-section global VAR model,} \emph{European Economic Review} \textbf{75},
  195--215.

\bibitem[{Gilbert(2010)}]{Gilbert2010}
\textsc{Gilbert CL} (2010), \enquote{{How to understand high food prices},}
  \emph{Journal of Agricultural Economics} \textbf{61}(2), 398--425.

\bibitem[{Griffin and Brown(2010)}]{griffin2010inference}
\textsc{Griffin JE, and Brown PJ} (2010), \enquote{Inference with normal-gamma
  prior distributions in regression problems,} \emph{Bayesian Analysis}
  \textbf{5}(1), 171--188.

\bibitem[{Guerrero \emph{et~al.}(2017)Guerrero, Hern{\'{a}}ndez-del Valle, and
  Ju{\'{a}}rez-Torres}]{Guerrero2017}
\textsc{Guerrero S, Hern{\'{a}}ndez-del Valle G, and Ju{\'{a}}rez-Torres M}
  (2017), \enquote{{Using a functional approach to test trending volatility in
  the price of Mexican and international agricultural products},}
  \emph{Agricultural Economics} \textbf{48}(1), 3--13.

\bibitem[{Harari and Ferrara(2018)}]{Harari2018}
\textsc{Harari M, and Ferrara EL} (2018), \enquote{{Conflict, climate and
  cells: A disaggregated analysis},} \emph{Review of Economics and Statistics}
  \textbf{100}(4), 594--608.

\bibitem[{Harri \emph{et~al.}(2009)Harri, Nalley, and Hudson}]{Harri2009}
\textsc{Harri A, Nalley L, and Hudson D} (2009), \enquote{{The Relationship
  between Oil, Exchange Rates, and Commodity Prices},} \emph{Journal of
  Agricultural and Applied Economics} \textbf{41}(02), 501--510.

\bibitem[{Havl{\'{i}}k \emph{et~al.}(2011)Havl{\'{i}}k, Schneider, Schmid,
  B{\"{o}}ttcher, Fritz, Skalsk{\'{y}}, Aoki, Cara, Kindermann, Kraxner, Leduc,
  McCallum, Mosnier, Sauer, and Obersteiner}]{Havlik2011}
\textsc{Havl{\'{i}}k P, Schneider Ua, Schmid E, B{\"{o}}ttcher H, Fritz S,
  Skalsk{\'{y}} R, Aoki K, Cara SD, Kindermann G, Kraxner F, Leduc S, McCallum
  I, Mosnier A, Sauer T, and Obersteiner M} (2011), \enquote{{Global land-use
  implications of first and second generation biofuel targets},} \emph{Energy
  Policy} \textbf{39}(1), 5690--5702.

\bibitem[{Headey(2011)}]{Headey2011}
\textsc{Headey D} (2011), \enquote{{Rethinking the global food crisis: The role
  of trade shocks},} \emph{Food Policy} \textbf{36}(2), 136--146.

\bibitem[{Huang \emph{et~al.}(2011)Huang, von Lampe, and van
  Tongeren}]{Huang2011}
\textsc{Huang H, von Lampe M, and van Tongeren F} (2011), \enquote{{Climate
  change and trade in agriculture},} \emph{Food Policy} \textbf{36}, S9--S13.

\bibitem[{Huber(2016)}]{huber2016density}
\textsc{Huber F} (2016), \enquote{Density forecasting using Bayesian global
  vector autoregressions with stochastic volatility,} \emph{International
  Journal of Forecasting} \textbf{32}(3), 818--837.

\bibitem[{Huber and Feldkircher(2019)}]{Huber2017}
\textsc{Huber F, and Feldkircher M} (2019), \enquote{Adaptive shrinkage in
  Bayesian vector autoregressive models,} \emph{Journal of Business \& Economic
  Statistics} \textbf{37}(1), 27--39.

\bibitem[{Huber \emph{et~al.}(2017)Huber, Krisztin, and Piribauer}]{Huber2016}
\textsc{Huber F, Krisztin T, and Piribauer P} (2017), \enquote{{Forecasting
  global equity indices using large Bayesian VARs},} \emph{Bulletin of Economic
  Research} \textbf{69}(3), 288--308.

\bibitem[{IFPRI(2008)}]{IFPRI2008}
\textsc{IFPRI} (2008), \emph{{High food prices: The what, who, and how of
  proposed policy actions}}, Washington DC: IFPRI.

\bibitem[{Ishwaran \emph{et~al.}(2001)Ishwaran, James, and
  Sun}]{ishwaran2001bayesian}
\textsc{Ishwaran H, James LF, and Sun J} (2001), \enquote{Bayesian model
  selection in finite mixtures by marginal density decompositions,}
  \emph{Journal of the American Statistical Association} \textbf{96}(456),
  1316--1332.

\bibitem[{Jaroci{\'n}ski(2010)}]{jarocinski2010responses}
\textsc{Jaroci{\'n}ski M} (2010), \enquote{Responses to monetary policy shocks
  in the east and the west of Europe: a comparison,} \emph{Journal of Applied
  Econometrics} \textbf{25}(5), 833--868.

\bibitem[{Jebabli \emph{et~al.}(2014)Jebabli, Arouri, and Teulon}]{Jebabli2014}
\textsc{Jebabli I, Arouri M, and Teulon F} (2014), \enquote{{On the effects of
  world stock market and oil price shocks on food prices: An empirical
  investigation based on TVP-VAR models with stochastic volatility},}
  \emph{Energy Economics} \textbf{45}, 66--98.

\bibitem[{Kastner(2019{\natexlab{a}})}]{Kastner2016}
\textsc{Kastner G} (2019{\natexlab{a}}), \enquote{Sparse Bayesian time-varying
  covariance estimation in many dimensions,} \emph{Journal of Econometrics}
  \textbf{210}(1), 98 -- 115.

\bibitem[{Kastner(2019{\natexlab{b}})}]{factorstochvol}
---{}---{}--- (2019{\natexlab{b}}), \emph{\texttt{factorstochvol}: {B}ayesian
  Estimation of (Sparse) Latent Factor Stochastic Volatility Models},
  \texttt{R}-package version 0.9.

\bibitem[{Kastner and Huber(2020)}]{kastner2017sparse}
\textsc{Kastner G, and Huber F} (2020), \enquote{Sparse Bayesian vector
  autoregressions in huge dimensions,} \emph{Journal of Forecasting}
  \textbf{39}(7), 1142--1165.

\bibitem[{Koop and Korobilis(2016)}]{koop2016model}
\textsc{Koop G, and Korobilis D} (2016), \enquote{Model uncertainty in panel
  vector autoregressive models,} \emph{European Economic Review} \textbf{81},
  115--131.

\bibitem[{Koop and Korobilis(2018)}]{koop2018forecasting}
---{}---{}--- (2018), \enquote{Forecasting with High-Dimensional Panel VARs,}
  \emph{Essex Finance Centre Working Papers} (31).

\bibitem[{Korobilis(2016)}]{korobilis2016prior}
\textsc{Korobilis D} (2016), \enquote{Prior selection for panel vector
  autoregressions,} \emph{Computational Statistics \& Data Analysis}
  \textbf{101}, 110--120.

\bibitem[{Leng and Hall(2019)}]{Leng2019}
\textsc{Leng G, and Hall J} (2019), \enquote{{Crop yield sensitivity of global
  major agricultural countries to droughts and the projected changes in the
  future},} \emph{Science of the Total Environment} \textbf{654}, 811--821.

\bibitem[{Lenk and DeSarbo(2000)}]{lenk2000bayesian}
\textsc{Lenk PJ, and DeSarbo WS} (2000), \enquote{Bayesian inference for finite
  mixtures of generalized linear models with random effects,}
  \emph{Psychometrika} \textbf{65}(1), 93--119.

\bibitem[{Lucotte(2016)}]{Lucotte2016}
\textsc{Lucotte Y} (2016), \enquote{{Co-movements between crude oil and food
  prices: A post-commodity boom perspective},} \emph{Economics Letters}
  \textbf{147}, 142--147.

\bibitem[{Malsiner-Walli \emph{et~al.}(2016)Malsiner-Walli,
  Fr{\"u}hwirth-Schnatter, and Gr{\"u}n}]{malsiner2016model}
\textsc{Malsiner-Walli G, Fr{\"u}hwirth-Schnatter S, and Gr{\"u}n B} (2016),
  \enquote{Model-based clustering based on sparse finite Gaussian mixtures,}
  \emph{Statistics and Computing} \textbf{26}(1-2), 303--324.

\bibitem[{Minot(2014)}]{Minot2014}
\textsc{Minot N} (2014), \enquote{{Food price volatility in sub-Saharan Africa:
  Has it really increased?}} \emph{Food Policy} \textbf{45}, 45--56.

\bibitem[{Nazlioglu(2011)}]{Nazlioglu2011a}
\textsc{Nazlioglu S} (2011), \enquote{{World oil and agricultural commodity
  prices: Evidence from nonlinear causality},} \emph{Energy Policy}
  \textbf{39}(5), 2935--2943.

\bibitem[{Nazlioglu and Soytas(2011)}]{Nazlioglu2011}
\textsc{Nazlioglu S, and Soytas U} (2011), \enquote{{World oil prices and
  agricultural commodity prices: Evidence from an emerging market},}
  \emph{Energy Economics} \textbf{33}(3), 488--496.

\bibitem[{Nazlioglu and Soytas(2012)}]{Nazlioglu2012}
---{}---{}--- (2012), \enquote{{Oil price, agricultural commodity prices, and
  the dollar: A panel cointegration and causality analysis},} \emph{Energy
  Economics} \textbf{34}(4), 1098--1104.

\bibitem[{Park and Casella(2008)}]{park2008bayesian}
\textsc{Park T, and Casella G} (2008), \enquote{The Bayesian Lasso,}
  \emph{Journal of the American Statistical Association} \textbf{103}(482),
  681--686.

\bibitem[{Pesaran \emph{et~al.}(2004)Pesaran, Schuermann, and
  Weiner}]{Pesaran2004}
\textsc{Pesaran MH, Schuermann T, and Weiner SM} (2004), \enquote{{Modeling
  regional interdependencies using a global error-correcting macroeconometric
  model},} \emph{Journal of Business and Economic Statistics} \textbf{22}(2),
  129--162.

\bibitem[{Pitt and Shephard(1999)}]{pitt1999time}
\textsc{Pitt M, and Shephard N} (1999), \enquote{Time varying covariances: a
  factor stochastic volatility approach,} \emph{Bayesian statistics}
  \textbf{6}, 547--570.

\bibitem[{Puma \emph{et~al.}(2015)Puma, Bose, Chon, and Cook}]{Puma2015}
\textsc{Puma MJ, Bose S, Chon SY, and Cook BI} (2015), \enquote{{Assessing the
  evolving fragility of the global food system},} \emph{Environmental Research
  Letters} \textbf{10}(2).

\bibitem[{Roberts and Schlenker(2013)}]{Roberts2010}
\textsc{Roberts MJ, and Schlenker W} (2013), \enquote{{Identifying supply and
  demand elasticities of agricultural commodities: Implications for the US
  ethanol mandate},} \emph{American Economic Review} \textbf{103}(6),
  2265--2295.

\bibitem[{Rousseau and Mengersen(2011)}]{rousseau2011asymptotic}
\textsc{Rousseau J, and Mengersen K} (2011), \enquote{Asymptotic behaviour of
  the posterior distribution in overfitted mixture models,} \emph{Journal of
  the Royal Statistical Society: Series B (Statistical Methodology)}
  \textbf{73}(5), 689--710.

\bibitem[{Saghaian(2010)}]{Saghaian2010}
\textsc{Saghaian SH} (2010), \enquote{{The Impact of the Oil Sector on
  Commodity Prices: Correlation or Causation?}} \emph{Journal of Agricultural
  and Applied Economics} \textbf{42}(03), 477--485.

\bibitem[{Sandstr{\"{o}}m \emph{et~al.}(2018)Sandstr{\"{o}}m, Valin, Krisztin,
  Havl{\'{i}}k, Herrero, and Kastner}]{Sandstrom2018}
\textsc{Sandstr{\"{o}}m V, Valin H, Krisztin T, Havl{\'{i}}k P, Herrero M, and
  Kastner T} (2018), \enquote{{The role of trade in the greenhouse gas
  footprints of EU diets},} \emph{Global Food Security} \textbf{19}, 48--55.

\bibitem[{Serra \emph{et~al.}(2011)Serra, Zilberman, Gil, and
  Goodwin}]{Serra2011}
\textsc{Serra T, Zilberman D, Gil JM, and Goodwin BK} (2011),
  \enquote{{Nonlinearities in the U.S. corn-ethanol-oil-gasoline price
  system},} \emph{Agricultural Economics} \textbf{42}(1), 35--45.

\bibitem[{van~der Velde \emph{et~al.}(2012)van~der Velde, Tubiello, Vrieling,
  and Bouraoui}]{VanderVelde2012}
\textsc{van~der Velde M, Tubiello FN, Vrieling A, and Bouraoui F} (2012),
  \enquote{{Impacts of extreme weather on wheat and maize in France: Evaluating
  regional crop simulations against observed data},} \emph{Climatic Change}
  \textbf{113}(3-4), 751--765.

\bibitem[{van Huellen(2018)}]{VanHuellen2018}
\textsc{van Huellen S} (2018), \enquote{{How financial investment distorts food
  prices: evidence from U.S. grain markets},} \emph{Agricultural Economics}
  \textbf{49}(2), 171--181.

\bibitem[{Yau and Holmes(2011)}]{yau2011hierarchical}
\textsc{Yau C, and Holmes C} (2011), \enquote{Hierarchical Bayesian
  nonparametric mixture models for clustering with variable relevance
  determination,} \emph{Bayesian Analysis} \textbf{6}(2), 329.

\end{thebibliography}

\newpage
\begin{center}
\begin{Large}
\textbf{Appendices}
\end{Large}
\end{center}
\begin{appendices}\crefalias{section}{appsec}
\setcounter{equation}{0}
\renewcommand\theequation{A.\arabic{equation}}
\section{Posterior simulation}\label{sec:posteriorsim}
\begin{enumerate}[align=left]
\item Simulation of VAR coefficients, factor loadin{}gs and stochastic volatility components
\begin{enumerate}[align=left]
\item Sample $\bm{A}_i$ and $\bm{B}_i$ from their Gaussian conditional posterior distributions on an equation-by-equation basis. Conditional on $\bm{L} \bm{f}_t$, the conditional posterior for each equation of \autoref{eq:PVAR} is given by
\begin{equation*}
\begin{pmatrix}
[\bm{C}_i]'_{j \bullet} \\
[\bm{B}_i]'_{j \bullet}
\end{pmatrix}|\bm{\bullet} \sim \mathcal{N}(\overline{\bm{c}}_{ij},\overline{\bm{M}}_{ij})
\end{equation*} 
for $i=1,\dots, N$ and $j=1,\dots, M$. The posterior mean and variance are given by
\begin{align*}
\overline{\bm{M}}_{ij} &= (\tilde{\bm{X}}'_i\tilde{\bm{X}}_i+\bm{W}_i^{-1})^{-1},\\
\overline{\bm{c}}_{ij} &= \overline{\bm{M}}_{ij} (\tilde{\bm{X}}'_i [\tilde{\bm{Y}}_i]_{\bullet j}+\bm{W}_{ij}^{-1} \bm{\psi}_{ij}),
\end{align*}
with $\tilde{\bm{X}}_i$ being a full-data matrix with typical $t$th row given by $(\bm{x}'_{it},\bm{x}'_{-i, t}) ~ \exp(-\omega_{tn}/2)$. The index  $n$ selects the  element of $\bm{\Omega}_t$ associated with the $j$th equation in country $i$ and $[\tilde{\bm{Y}}_i]_{\bullet j}$ has typical element ${y}_{ij, t} - [\bm{L}]_{n \bullet} \bm{f}_t$. In addition, $\bm{W}_i= \text{diag}(\bm{V}_j, \bm{\Delta}_{ij})$ with $\bm{\Delta}_{ij}$ being a diagonal prior variance-covariance matrix for the $j$th equation constructed using \autoref{eq:NG_int}, and $\bm{\psi}_{ij}$ is a prior mean matrix that consists of the elements in $\bm{\mu}_g$ associated with the $j$th equation, for $\delta_i =g$, and the remaining elements are set equal to zero. The matrix $\bm{V}_j$ is constructed by selecting the variance parameters in $\bm{V}$ that relate to the $j$th equation.

\item We simulate the quantities related to the factor stochastic volatility specification using the \texttt{R}-package \texttt{factorstochvol} \citep{factorstochvol}. Details on the posterior quantities involved are presented in \citet{Kastner2016}. Specifically, the algorithm draws first the full history of the idiosyncratic volatilities $h_{jt}$, the volatilities of the factors $\omega_{jt}$ and the corresponding parameters $\phi_{\omega j}$, $\rho_{sj}$ and $\sigma_{sj}$. The subsequent step produces draws for the global and local shrinkage parameters from the row-wise Normal-Gamma prior specification. In the next step, simple Bayesian regression updates can be used to obtain a draw for the factor loadings matrix $\bm{L}$. To speed up mixing in this step, the algorithm relies on deep interweaving techniques. The final updating step produces a draw for the full history of the factors $\bm{f}_t$.
\end{enumerate}

\item Simulation of quantities associated with the mixture model
\begin{enumerate}[align=left]
\item Sample the mixture probabilities $\bm{w}$ from a Dirichlet distribution given by
\begin{equation*}
\bm{w}|\bm{\bullet} \sim \textit{Dir}(p_1, \dots, p_G),
\end{equation*}
with $p_g = p_0 + N_g$ and $N_g=\#\{i: \delta_i = g\}$ denoting the number of countries located in cluster $g$. 
\item The regime indicators $\delta_i$ are simulated from a multinomial distribution with
\begin{equation*}
\Pr(\delta_i = k ) \propto w_k f_\mathcal{N}(\bm{c}_i | \bm{\mu}_g, \bm{V}).
\end{equation*}

\item We obtain draws for the group-specific means from a multivariate Gaussian distribution,
\begin{align*}
&\bm{\mu}_g | \bm{\bullet} \sim \mathcal{N}(\overline{\bm{\mu}}_g, \overline{\bm{V}}_g),\\
&\overline{\bm{V}}_g =  \left(N_g \bm{V}^{-1}+ \bm{Q}_0^{-1}\right)^{-1},\\
&\overline{\bm{\mu}}_g =\overline{\bm{V}}_g \left(N_g \bm{V}^{-1}\overline{\bm{c}}_g + \bm{Q}_0^{-1}\bm{\mu}_0 \right).
\end{align*}
$\overline{\bm{c}}_g=\frac{\sum_{i=1}^N \bm{c}_i \delta_i}{N_g}$ denotes the mean of the domestic quantities associated with group $g$.

\item The common variance-covariance matrix $\bm{V}$ is obtained by independently sampling $v_j~(j=1,\dots,m)$ from 
\begin{equation*}
v_j|\bm{\bullet} \sim \mathcal{G}^{-1}\left(w_0+\frac{N}{2}, w_1 + \frac{\sum_{n=1}^N (c_{nj} - \mu_{nj})^2}{2}\right),
\end{equation*}
where $\mu_{nj}=\mu_{gj}$ if $\delta_n=g$. 

\item We simulate $\lambda_j$ from a generalized inverted Gaussian (GIG) distribution,\footnote{We assume that $x$ follows a GIG distribution if its density is proportional to  $x^{a-1} \exp\{-(bx + c/x)/2\}$ with $a \in \mathbb{R}$ and  $b, c>0$. }
\begin{equation*}
\lambda_j | \bm{\bullet} \sim \mathcal{GIG}(p_G, d_j, e_j).
\end{equation*}
After simulating all $\lambda_j$s  we construct $\bm{Q}_0 = \bm{\Lambda} \bm{R}_0 \bm{\Lambda}$, with $\bm{R}_0$ being based on the most recent Gibbs draw of $\bm{c}$.
 \item The full conditional posterior of $\bm{\mu}_0$ is Gaussian with
 \begin{equation*}
 \bm{\mu}_0|\bm{\bullet} \sim \mathcal{N}(\overline{\bm{\mu}}_0,  \overline{\bm{Q}}_0),
 \end{equation*}
whereby $\overline{\bm{\mu}}_0 = \frac{1}{G} \sum_{g=1}^G \bm{\mu}_g$ and $\overline{\bm{Q}}_0=\frac{1}{G} \bm{Q}_0$.

\item Simulate the intensity parameter of the Dirichlet prior $p_0$ using a random walk Metropolis Hastings algorithm  on the log scale. The full conditional posterior density of $p_0$ is given by
\begin{equation*}
p(p_0| \bm{w}) \propto p(\bm{w}|p_0)~ p(p_0). \label{eq: RWMH}
\end{equation*}
We propose a value  $p^*_0$  from $ p^*_0 \sim p^{(a)}_0 e^{z}$ with $z\sim \mathcal{N}(0, \mathfrak{c})$. Here we let $\mathfrak{c}$ be a tuning parameter specified such that the acceptance rate lies between 20 and 40 percent and $p^{(a)}_0$ denotes the last accepted draw. The probability of accepting a new draw is then
\begin{equation*}
\alpha(p^*_0, p^{(a)}_0) = \text{min}\left[\frac{p(\bm{w}|p^*_0)~ p(p^*_0) ~p^*_0}{p(\bm{w}|p^{(a)}_0)~ p(p^{(a)}_0) ~p^{(a)}_0}, 1\right].
\end{equation*}

\end{enumerate}
\item Simulation of shrinkage parameters on dynamic interdependencies
\begin{enumerate}[align=left]
\item For each country $i=1,\dots,N$, simulate the global shrinkage parameters $\xi_i$ from a Gamma distribution,
\begin{equation*}
\xi_i | \bm{\bullet} \sim \mathcal{G}\left(\mathfrak{c}_0+\vartheta_i k, \mathfrak{c}_0 + \frac{\vartheta_i}{2}\sum_{i=1}^k \tau_{ij}\right).
\end{equation*}
\item Sample the local shrinkage parameters from their GIG distributed posteriors 
\begin{equation*}
\tau_{ij}|\bm{\bullet} \sim \mathcal{GIG}\left(\vartheta_i - \frac{1}{2}, \vartheta_i \xi_i, b_{ij}^2\right),
\end{equation*}
for $i=1,\dots, N$ and $j=1,\dots, k$.
\end{enumerate}
\item Apply a random permutation step by simulating one of $G!$ possible permutations of $\{1, \dots, G\}$, labeled  $\varrho$,
\begin{align*}
&(w_1, \dots, w_G)' = (w_{\varrho(1)},\dots, w_{\varrho(G)}),\\
&(\bm{\mu}_1, \dots, \bm{\mu}_G)' = (\bm{\mu}_{\varrho(1)},\dots, \bm{\mu}_{\varrho(G)}),\\
&\bm{\delta} = \varrho(\bm{\delta}).
\end{align*}
\end{enumerate}

\setcounter{equation}{0}
\renewcommand\theequation{B.\arabic{equation}}
\section{Simulation based evidence}\label{sec: simulation}
In this section we evaluate the merits of our approach by means of an extensive simulation exercise. To this end, we consider a range of alternative DGPs and scenarios which differ in terms of the implied sparsity on the dynamic interdependencies in $\bm{B}_i$ as well as the length of the time period. 

The DGP we consider assumes that $N=26, M=2, P=1$ and $T \in \{80, 150, 250\}$. The $M$ variables per country are labeled $UN$ and $DP$. Here, and in  the empirical application, these acronyms refer to unemployment and inflation, respectively.  Moreover, we assume that the domestic coefficients (including the intercept) come from a two component mixture of Gaussians (i.e. $G^{true}=2$) with mean vectors given by $\bm{\mu}^{true}_1=(0.6, 0.2,2.0, 0.3, 0.6,-3.0)'$, $\bm{\mu}^{true}_2=(-0.6, 0.2,5, -0.8, 0.6,0.0)'$,  variance-covariance matrix $\bm{V}^{true}=\frac{1}{10^3}\times \bm{I}$ and $\bm{w}^{true}= (0.4, 0.6)'$.  Notice that across clusters, the coefficients associated with lagged DP in the UN  equations as well as the first, own lag of DP are equal across clusters. This serves as a simple test whether the NG shrinkage prior successfully shrinks the corresponding differences in cluster centers to zero.

Coefficients measuring lagged interdependencies are constructed  by drawing from univariate Gaussian distributions $b^{true}_{ij} \sim \mathcal{N}\left(0, \frac{1}{10^2}\right)$. To control the degree of sparsity, we zero out all coefficients that are below a fraction, denoted by $\varpi \in \{0.15, 0.30, 0.60, 0.90\}$, of the maximum absolute value across all $b_{ij}$'s. Finally, for the factor model in \autoref{eq: factorSV}, we use $q=2$, sample the loadings from zero mean Gaussian distributions with variances given by $1/(1000^2)$, specify $\rho_{hj}=0.9$ and $\sigma_{hj}=0.1$ for all $j$, set $\phi_{\omega j}=-10, \rho_{\omega j}=0.9$ and $\sigma_{\omega j}=0.01$. For each of the different DGPs, we run a set of $50$ simulations and focus on root mean square errors (RMSEs).

Before showing the results, a brief word on the specification and identification of the underlying mixture model is necessary.   We assume that $G=10$, implying that $G \gg G^{true}$, the number of factors equals  the true number $q=2$, and  set the lag length to $P=1$. In this simulation exercise, we identify the model by applying the permutation sampler and identify the mixture model by assuming that $w_1 < \dots <w_G$. The hyperparameters are specified as described in Section \ref{sec: econometrics}.

As competing alternatives, we include a model estimated with a single regime ($Q=1$),  a flat prior VAR (labeled VAR-OLS), and a VAR with a NG shrinkage prior (VAR-NG) that treats the PVAR as a large VAR without discriminating, a priori, between domestic and foreign variables. All models, except the flat prior VAR, feature a factor stochastic volatility specification in the reduced form errors.  For the  VAR-NG model, we use the lag-wise prior specification described in \cite{Huber2017} with the choice of the hyperparameters closely mirroring the ones used  on the DIs.

\begin{table*}[ht] 
\caption{Root mean square errors of posterior median of the VAR coefficients and the true values.}\vspace*{-1.8em}
\footnotesize
\begin{center}
\begin{threeparttable}
\begin{tabular}{lrrrrrrrrrrrr}
\toprule
 & \multicolumn{4}{c}{$T=80$}&\multicolumn{4}{c}{$T=150$}&\multicolumn{3}{c}{$T=250$} \\
 \cmidrule(l{0pt}r{3pt}){2-5} \cmidrule(l{3pt}r{3pt}){6-9} \cmidrule(l{3pt}r{0pt}){10-13}
 Sparsity& 0.15 & 0.30 & 0.60 & 0.90& 0.15 & 0.30 & 0.60 & 0.90 & 0.15 & 0.30 & 0.60 & 0.90 \\
\midrule
PVAR $G=8$  & 0.026 & 0.026 & 0.025 & 0.024 & 0.021 & 0.021 & 0.019 & 0.019 & 0.017 & 0.016 & 0.015 & 0.014 \\
  PVAR $G=1$& 0.031 & 0.031 & 0.030 & 0.030 & 0.025 & 0.025 & 0.023 & 0.023 & 0.019 & 0.019 & 0.018 & 0.017 \\
  VAR-NG  & 0.042 & 0.041 & 0.041 & 0.041 & 0.031 & 0.031 & 0.030 & 0.030 & 0.022 & 0.022 & 0.021 & 0.020 \\ 
  VAR-OLS   & 0.179 & 0.177 & 0.179 & 0.178 & 0.102 & 0.100 & 0.101 & 0.101 & 0.061 & 0.061 & 0.060 & 0.060 \\
   \bottomrule
\end{tabular}
\begin{tablenotes}[para,flushleft]
\footnotesize{\textit{Notes}: Sparsity refers to the zeroing out all coefficients that are below the indicated fraction of the maximum absolute value across all dynamic interdependencies. Values represent the median across $50$ replications.}
\end{tablenotes}
\end{threeparttable}
\end{center}
\label{tab: simulation1}
\end{table*}

Table \ref{tab: simulation1}  shows the median across replications of the RMSEs for the different models. A few findings are worth emphasizing.  First, notice that irrespective of the DGP adopted, our proposed overfitting mixture model performs particularly well. In fact, it  outperforms all competing models considered. The misspecified model with $G=1$ also yields relatively precise estimates. Unsurprisingly, the VAR estimated by OLS performs worst. Second, considering different lengths of the sample reveals that if more information is available, the accuracy gap between the unrestricted VARs (i.e. estimated using OLS and under the lag-wise NG prior) and the two PVAR specifications tends to decline.  Nevertheless, OLS still yields estimation errors that are sizable and about four times as large as the estimation errors obtained by using the PVAR with $G=8$ (for $T=250$). From a practical perspective, and by considering the unrestricted VAR with the NG prior, this simply implies that for large $T$,  simpler specifications also tend to work well.

Third, and finally, we observe a better performance of our proposed framework  for higher levels of sparsity, especially when benchmarked against unrestricted specifications. This finding indicates that separately controlling for dynamic interdependencies evidently improves model performance as compared to a prior that does not discriminate between coefficients associated with foreign quantities and their domestic counterparts.

To assess whether our mixture model successfully detects the correct number of regimes as well as the correct regime allocation, Table \ref{tab: sim2} shows the (half) quadratic probability score (QPS) defined as,
\begin{equation}
\text{QPS}= \frac{1}{N}\sum_{i=1}^N (\delta^{true}_i- \overline{\delta}_i)^2,
\end{equation}
with $\delta^{true}_i$ denoting the true cluster allocation for country $i$ and $\overline{\delta}_i$ is the posterior mean of $\overline{\delta}_i$. All numbers in the table refer to the median across simulation replications. The QPS score is bounded between zero and unity, with a value of zero indicating perfect accuracy.

The results in Table \ref{tab: sim2} suggest that irrespective of the DGP,  our model appears to work quite well, yielding QPS scores close to zero.  Interestingly, differences in QPS scores tend to be quite unsystematic, providing only limited evidence that accuracy improves if the length of the sample is increased (see the final two rows of the table). 
This finding can be traced back to the fact that our DGP induces quite large differences between $\bm{\mu}^{true}_1$ and $\bm{\mu}^{true}_2$, implying that the conditional likelihood carries sufficient information (see \autoref{eq: mixtures}). In addition, notice that  the number of countries is the same across DGPs, implying that the number of observations from which we infer the correct clustering stays the same. The robustness of the QPS scores with respect to the sparsity level implies that the actual level of sparsity on DIs does not exert a significant feedback effect on the actual cluster allocation.

\begin{table*}
\caption{Quadratic probability score across simulation runs.}\vspace*{-1.8em}
\footnotesize
\begin{center}
\begin{threeparttable}
\begin{tabular*}{0.8\textwidth}{@{\extracolsep{\fill}}lccc}
  \toprule
   Sparsity   & $T=80$ & $T=150$ & $T=250$ \\ 
  \midrule
  0.15  & 0.155 & 0.124 & 0.156 \\  
  0.3   & 0.124 & 0.156 & 0.155 \\    
  0.6   & 0.156 & 0.123 & 0.126 \\    
  0.9   & 0.156 & 0.154 & 0.140 \\    
   \bottomrule
\end{tabular*}
\begin{tablenotes}[para,flushleft]
\footnotesize{\textit{Notes}: Sparsity refers to the zeroing out all coefficients that are below the indicated fraction of the maximum absolute value across all dynamic interdependencies.}
\end{tablenotes}
\end{threeparttable}
\end{center}
\label{tab: sim2}
\end{table*}

\begin{figure}[t]
\begin{center}
\includegraphics[scale=.5]{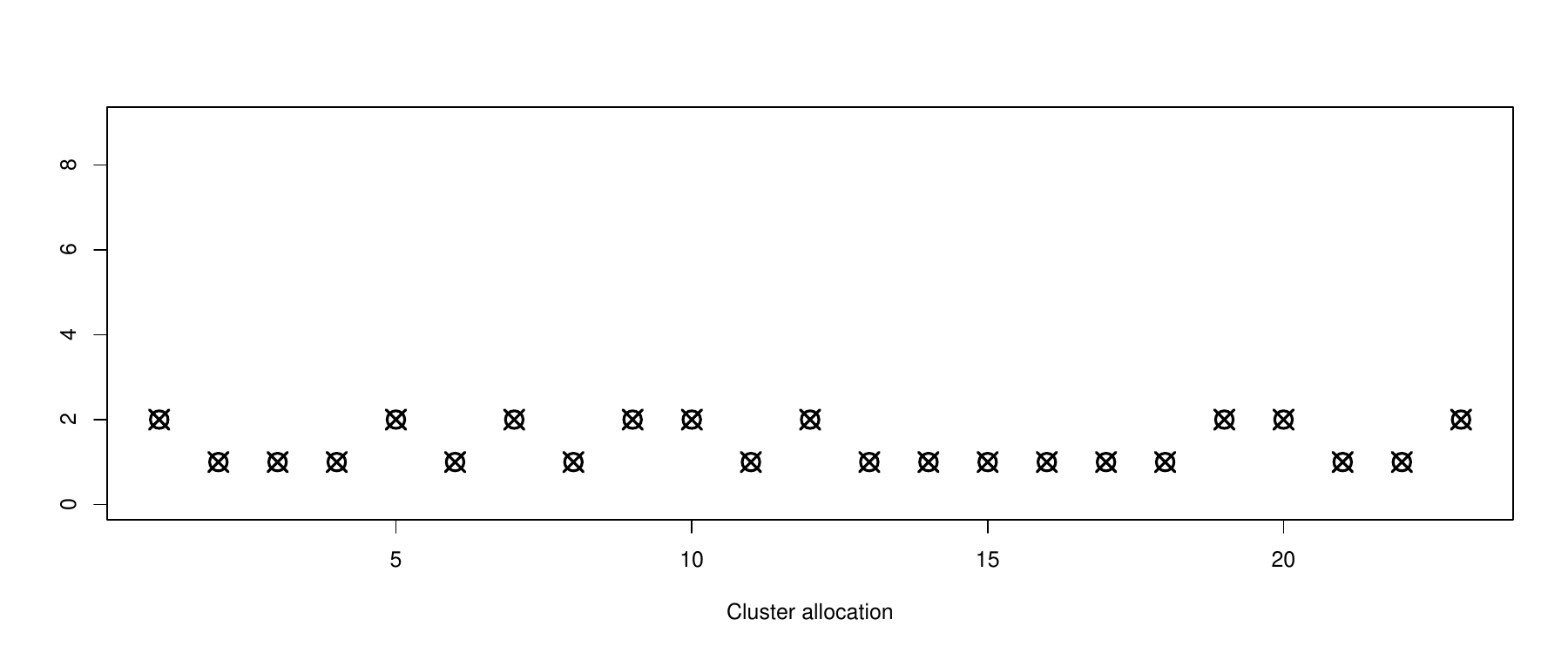}
\end{center}\vspace*{-2em}
\caption{Estimated regime allocation across countries and true allocation.}
\footnotesize\textit{Notes}: Black circles refer to the estimated regime, while the true allocation is indicated by the black crosses. The vertical axis indicates cluster membership of each country, and the horizontal axis denotes the respective cross-sectional unit.
\label{fig: regimes_sim}
\end{figure}

After providing some evidence that our model performs well, we proceed by showing selected empirical features of the proposed framework for a single simulation run.   In Fig. \ref{fig: regimes_sim}, we show the estimated regime allocation across countries alongside the true regime allocation. Black circles represent the posterior mean of the regime indicators while black crosses mark the actual country allocation. Consistent with the findings reported in \autoref{tab: sim2}, the figure suggests that our model performs well in detecting country clusters and selecting the true number of regimes.

Figure \ref{fig: regimes_sim} shows the posterior distribution of $\bm{c}_i$ for all $i$ as well as the posterior density of the cluster centers $\bm{\mu}_j$ for $j=1,2$. For the country-specific coefficients we use red colored densities to indicate group membership of a country  within a group, blue densities point towards membership in other groups and the black density is the corresponding element of $\bm{\mu}_g$. In the figure titles, we first state the equation (i.e. the equation for UN or DP) and afterwards the variable within each equation, with $\beta$ referring to the intercept.

\begin{figure}[t]
\begin{subfigure}{.5\textwidth}
\includegraphics[scale=0.45]{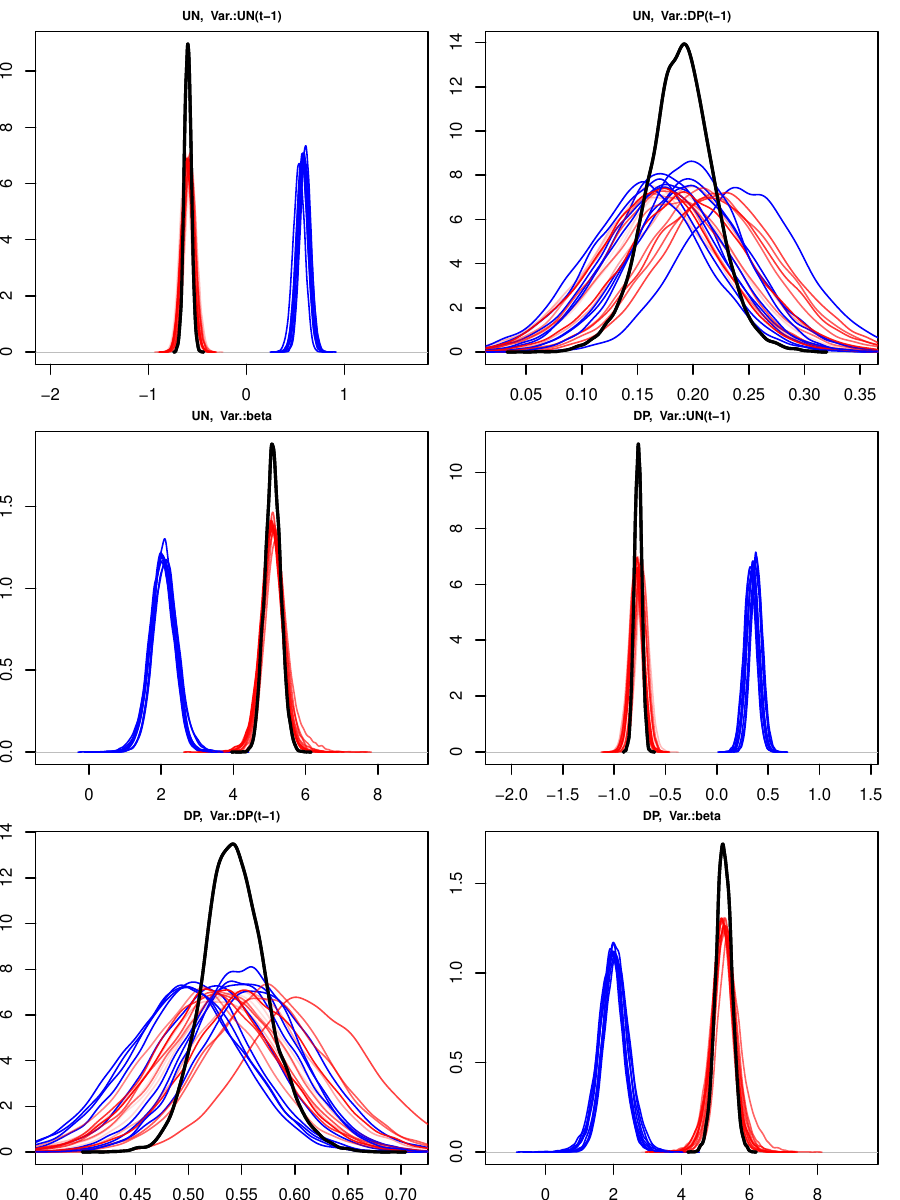}
\caption{Posterior distribution of $\bm{\mu}_1$}
\end{subfigure}
\begin{subfigure}{.5\textwidth}
\includegraphics[scale=0.45]{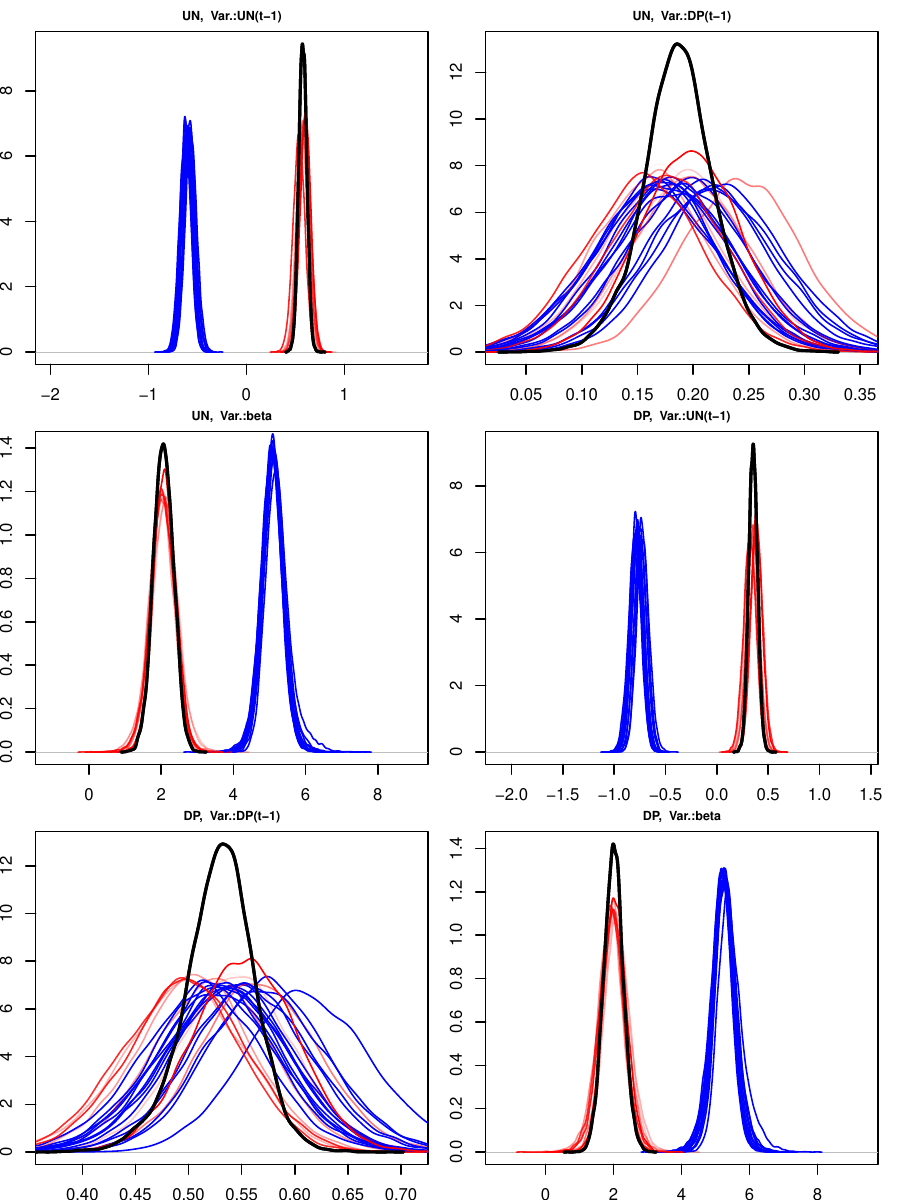}\caption{Posterior distribution of $\bm{\mu}_2$}
\end{subfigure}
\caption{Posterior distribution of $\bm{c}_i$ for simulated data.}
\footnotesize\textit{Notes}: The group specific mean $\bm{\mu}_i$ is depicted in solid black. Red lines indicate that a given country is member of the group considered while blue lines indicate membership to other groups.
\label{fig:dens_groups_sim}
\end{figure}
\begin{figure}[!ht]
\begin{center}
\includegraphics[scale=.4]{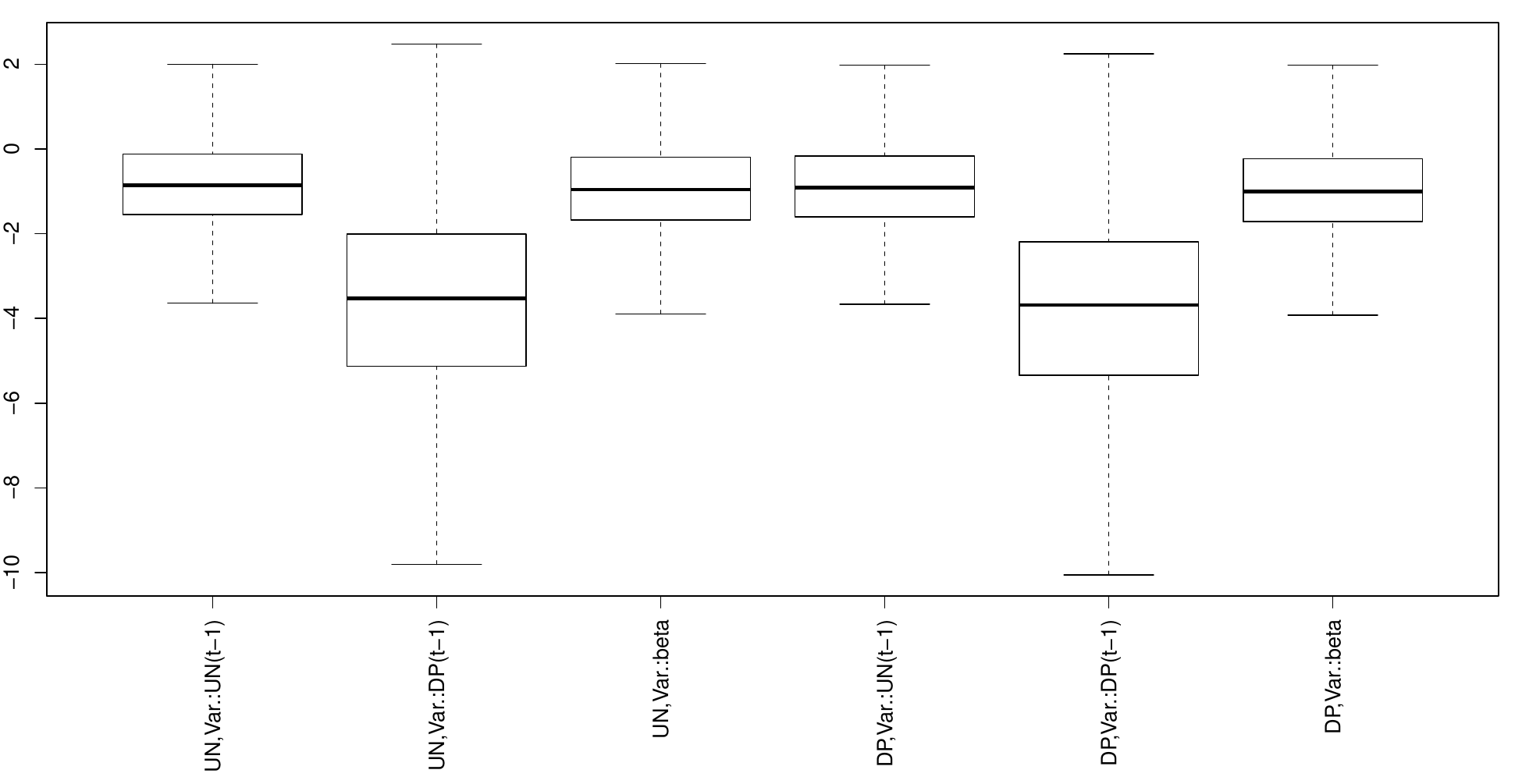}
\end{center}
\caption{Posterior distribution of $\log(\lambda_j)$ for the synthetic dataset}\label{fig: log_lambda}
\end{figure}

For each coefficient, the country-specific posterior distributions are closely centered on the posterior distribution of the group-specific coefficient. This holds true for both clusters that are not emptied out in MCMC sampling. Considering the coefficients that are homogenous across both clusters reveals that our model approach successfully detects homogeneity, pushing the country-specific estimates on lagged inflation  towards the common mean. Comparing the posterior densities across regimes for these coefficients also shows that our shrinkage prior successfully pulls the standardized distance between cluster centers to zero. Following \cite{yau2011hierarchical}, this finding is corroborated by considering boxplots of the log posterior distribution  of $\lambda_j$, depicted in \autoref{fig: log_lambda}. Here, we observe that for the coefficients specified to be equal across clusters, the corresponding (log) shrinkage factor is much smaller.

\FloatBarrier

\clearpage\setcounter{equation}{0}
\renewcommand\theequation{C.\arabic{equation}}
\section{Regions and data}\label{sec: app_regdat}

\begin{table*}[ht]
\caption{Composition of climatic regions.}\vspace*{-0.8em}
\tiny
\begin{tabular*}{\textwidth}{@{\extracolsep{\fill}}llll}
\toprule
\multicolumn{1}{c}{\textbf{Canada (CAN)}} & \multicolumn{1}{c}{\textbf{Europe (EUR)}} & \multicolumn{1}{c}{\textbf{Middle-East and North-Africa (MNA)}} & \multicolumn{1}{c}{\textbf{Sub-Saharan Africa (SSA)}} \\
Canada & Romania & Lebanon & \multicolumn{1}{l}{Burundi} \\
\multicolumn{1}{c}{\textbf{Russia and West Asia (CSI)}} & Serbia and Montenegro & Libya & \multicolumn{1}{l}{Cameroon} \\
Armenia & Slovakia & Morocco & \multicolumn{1}{l}{Cape Verde} \\
Azerbaijan & Slovenia & Oman & \multicolumn{1}{l}{Central African Republic} \\
Belarus & Spain & Qatar & \multicolumn{1}{l}{Chad} \\
Georgia & Sweden & Saudi Arabia & \multicolumn{1}{l}{Comoros} \\
Kazakhstan & Switzerland & Syria & \multicolumn{1}{l}{Congo} \\
Kyrgyzstan & United Kingdom & Tunisia & \multicolumn{1}{l}{Congo DR} \\
Moldova & \multicolumn{1}{c}{\textbf{Latin America (LAC)}} & Turkey & \multicolumn{1}{l}{Cote d'Ivoire} \\
Russian Federation & Argentina & United Arab Emirates & \multicolumn{1}{l}{Djibouti} \\
Tajikistan & Bahamas & Western Sahara & \multicolumn{1}{l}{Equatorial Guinea} \\
Turkmenistan & Belize & Yemen & \multicolumn{1}{l}{Eritrea} \\
Ukraine & Bolivia & \multicolumn{1}{c}{\textbf{Oceania (OCE)}} & \multicolumn{1}{l}{Ethiopia} \\
Uzbekistan & Brazil & Australia & \multicolumn{1}{l}{Gabon} \\
\multicolumn{1}{c}{\textbf{East Asia (EAS)}} & Chile & Fiji & \multicolumn{1}{l}{Gambia} \\
China & Colombia & French Polynesia & \multicolumn{1}{l}{Ghana} \\
Japan & Costa Rica & New Caledonia & \multicolumn{1}{l}{Guinea} \\
Republic of Korea & Cuba & New Zealand & \multicolumn{1}{l}{Guinea-Bissau} \\
\multicolumn{1}{c}{\textbf{Europe (EUR)}} & Dominican Republic & Papua New Guinea & \multicolumn{1}{l}{Kenya} \\
Albania & Ecuador & Samoa & \multicolumn{1}{l}{Lesotho} \\
Austria & El Salvador & Solomon Islands & \multicolumn{1}{l}{Liberia} \\
Belgium & Falkland Islands & Vanuatu & \multicolumn{1}{l}{Madagascar} \\
Bosnia and Herzegovina & French Guiana & \multicolumn{1}{c}{\textbf{South Asia (SAS)}} & \multicolumn{1}{l}{Malawi} \\
Bulgaria & Guadeloupe & Bangladesh & \multicolumn{1}{l}{Mali} \\
Croatia & Guatemala & Bhutan & \multicolumn{1}{l}{Mauritania} \\
Cyprus & Guyana & India & \multicolumn{1}{l}{Mauritius} \\
Czech Republic & Haiti & Nepal & \multicolumn{1}{l}{Mozambique} \\
Denmark & Honduras & Pakistan & \multicolumn{1}{l}{Namibia} \\
Estonia & Jamaica & Sri Lanka & \multicolumn{1}{l}{Niger} \\
Finland & Mexico & \multicolumn{1}{c}{\textbf{Southeast Asia (SEA)}} & \multicolumn{1}{l}{Nigeria} \\
France & Nicaragua & Brunei Darussalam & \multicolumn{1}{l}{Reunion} \\
Germany & Panama & Cambodia & \multicolumn{1}{l}{Rwanda} \\
Greece & Paraguay & Indonesia & \multicolumn{1}{l}{Senegal} \\
Greenland & Peru & Korea DPR & \multicolumn{1}{l}{Sierra Leone} \\
Hungary & Suriname & Laos & \multicolumn{1}{l}{Somalia} \\
Iceland & Trinidad and Tobago & Malaysia & \multicolumn{1}{l}{South Africa} \\
Ireland & Uruguay & Mongolia & \multicolumn{1}{l}{Sudan} \\
Italy & Venezuela & Myanmar & \multicolumn{1}{l}{Swaziland} \\
Latvia & \multicolumn{1}{c}{\textbf{Middle-East and North-Africa (MNA)}} & Philippines & \multicolumn{1}{l}{Tanzania} \\
Lithuania & Algeria & Singapore & \multicolumn{1}{l}{Togo} \\
Luxembourg & Bahrain & Thailand & \multicolumn{1}{l}{Uganda} \\
Macedonia & Egypt & Viet Nam & \multicolumn{1}{l}{Zambia} \\
Malta & Iran & \multicolumn{1}{c}{\textbf{Sub-Saharan Africa (SSA)}} & \multicolumn{1}{l}{Zimbabwe} \\
Netherlands & Iraq & Angola & \multicolumn{1}{c}{\textbf{United States of America (USA)}} \\
Norway & Israel & Benin & \multicolumn{1}{l}{United States of America} \\
Poland & Jordan & Botswana &  \\
Portugal & Kuwait & Burkina Faso &  \\
\bottomrule
\end{tabular*}
\begin{tablenotes}[para,flushleft]
\tiny{\textit{Notes}: Aggregated global climatic regions and the associated countries.}
\end{tablenotes}
\label{tab:Appendix_regions}
\end{table*}

\begin{table}[ht] 
\caption{Variables used in the PVAR model.}\vspace*{-2em}
\footnotesize
\begin{center}
\begin{tabular*}{\textwidth}{@{\extracolsep{\fill}}ll}
  \toprule
\textbf{Futures prices} & \textbf{Description and source} \\
(Abbreviation) & \\
 \midrule
 		 crude.oil & Intercontinental Exchange Brent Crude Futures, Continuous Contract \#2 \\
		  & (Source: \textit{Quandl/CHRIS}) \\
          gasoil & Intercontinental Exchange Gas Oil Futures, Continuous Contract \#2 \\
          & (Source: \textit{Quandl/CHRIS})  \\
          corn & Chicago Mercantile Exchange Corn Futures, Continuous Contract \#2 \\
          & (Source: \textit{Quandl/CHRIS})  \\
          rice & Chicago Mercantile Exchange Rough Rice Futures, Continuous Contract \#2 \\
          & (Source: \textit{Quandl/CHRIS}) \\
          soya & Chicago Mercantile Exchange Soya Futures, Continuous Contract \#2 \\
          & (Source: \textit{Quandl/CHRIS}) \\
          soyoil & Chicago Mercantile Exchange Soybean Oil Futures, Continuous Contract \#2 \\
          & (Source: \textit{Quandl/CHRIS}) \\
          wheat & Chicago Mercantile Exchange Wheat Futures, Continuous Contract \#2 \\
          & (Source: \textit{Quandl/CHRIS}) \\
          cotton & Intercontinental Exchange Cotton No. 2 Futures, Continuous Contract \#2 \\
          & (Source: \textit{Quandl/CHRIS}) \\
          ethanol & Chicago Mercantile Exchange Ethanol Futures, Continuous Contract \#2 \\
          & (Source: \textit{Quandl/CHRIS}) \\
          hogs & Chicago Mercantile Exchange Lean Hogs Futures, Continuous Contract \#2 \\
          & (Source: \textit{Quandl/CHRIS}) \\
          fcattle & Chicago Mercantile Exchange Feeder Cattle Futures, Continuous Contract \#2 \\
          & (Source: \textit{Quandl/CHRIS}) \\
          lcattle & Chicago Mercantile Exchange Live Cattle Futures, Continuous Contract \#2 \\
          & (Source: \textit{Quandl/CHRIS}) \\
 \midrule
 \textbf{Macroeconomic variables} & \textbf{Description and source} \\
(Abbreviation) & \\
 \midrule
		  CPI & Nominal seasonally adjusted consumer price index, measured in log differences. \\
 		  & (Source: \textit{World Bank})\\
          CPF & Ratio of nominal, seasonally adjusted food consumer price index and CPI, measured in \\
          & log levels. (Source: \textit{FAO})\\
          ir & Short term interest rates, percent per annum, measured in log levels. \\
          & (Source: \textit{OECD}) \\
          prod & Total value of industrial production, in seasonally adjusted 2010 USD, measured in \\
          & log levels. (Source \textit{World Bank})\\
          XR2 & Real effective exchange rate index, measured in log levels. \\
          & (Source: \textit{World Bank}) \\
   \bottomrule
\end{tabular*}
\begin{tablenotes}[para,flushleft]
\footnotesize{\textit{Notes}: Abbreviations and variable descriptions alongside the corresponding source of the data.}
\end{tablenotes}
\label{tab:Appendix_vars}
\end{center}
\end{table}
\end{appendices}

\end{document}